\documentclass[fleqn,usenatbib]{mnras}

\usepackage{newtxtext,newtxmath}

\usepackage[T1]{fontenc}

\DeclareRobustCommand{\VAN}[3]{#2}
\let\VANthebibliography\thebibliography
\def\thebibliography{\DeclareRobustCommand{\VAN}[3]{##3}\VANthebibliography}

\usepackage{graphicx}	
\usepackage{amsmath}	

\usepackage{threeparttable}
\usepackage[dvipsnames]{xcolor}
\usepackage{rotating} 

\newcommand{\Rs}{$\mathcal{R}_s$}

\newcommand{\cone}{[C\,{\footnotesize I}]}
\newcommand{\ctwo}{[C\,{\footnotesize II}]}

\newcommand{\as}{$^{\prime\prime}$}

\newcommand{\x}{$\times$}

\newcommand{\oii}{[O\,{\footnotesize II}]}

\newcommand{\cione}{[C\,{\footnotesize I}]$\,(^3P_1\,-\,^{3}P_0)$}
\newcommand{\citwo}{[C\,{\footnotesize I}]$\,(^3P_2\,-\, ^{3}P_1)$}

\newcommand{\cofour}{CO\,$(4-3)$}
\newcommand{\cothree}{CO\,$(3-2)$}
\newcommand{\cotwo}{CO\,$(2-1)$}
\newcommand{\coone}{CO\,$(1-0)$}

\newcommand{\hst}{\textit{HST}}
\newcommand{\spitzer}{\textit{Spitzer}}
\newcommand{\um}{$\mu$m}

\definecolor{purple}{rgb}{0.6, 0.4, 0.8}

\newcommand{\orcid}[1]{\textsuperscript{\,\,\href{https://orcid.org/#1}{\includegraphics[scale=0.06]{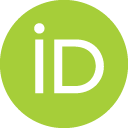}}}}

\title[Dust rich quiescent galaxy at $z=2$]{High dust content of a quiescent galaxy at $z\sim2$ revealed by deep ALMA observation}

\author[Minju. M. Lee et al.]{Minju M. Lee$^{1,2}$\orcid{0000-0002-2419-3068},
Charles C. Steidel$^{3}$\orcid{0000-0002-4834-7260},
Gabriel Brammer$^{1,4}$,
Natascha F\"{o}rster-Schreiber$^{5}$, Alvio Renzini$^{6}$,
\newauthor{Daizhong Liu$^{5}$, Rodrigo Herrera-Camus$^{7}$, Thorsten Naab$^{8}$, Sedona H. Price$^{9}$, Hannah \"{U}bler$^{10,11}$\orcid{0000-0003-4891-0794}, Sebasti\'{a}n }
\newauthor{Arriagada$^{7}$, and Georgios Magdis$^{1,2,4}$}\\
$^{1}$Cosmic Dawn Center (DAWN), Denmark \\
$^{2}$DTU-Space, Technical University of Denmark, Elektrovej 327, DK2800 Kgs. Lyngby, Denmark\\
$^{3}$Cahill Center for Astronomy and Astrophysics, California Institute of Technology, MS249-17, Pasadena, CA 91125, USA\\
$^{4}$Niels Bohr Institute, University of Copenhagen, Jagtvej 128, DK2200 Copenhagen N, Denmark\\
$^{5}$Max-Planck-Institut für Extraterrestrische Physik (MPE), Giessenbachstr. 1, D-85748 Garching, Germany\\
$^{6}$Osservatorio Astronomico di Padova, Vicolo dell'Osservatorio 5, Padova, I-35122, Italy\\
$^{7}$Astronomy Department, Universidad de Concepci\'{o}n, Av. Esteban Iturra s/n Barrio Universitario, Casilla 160, Concepci\'{o}n, Chile\\
$^{8}$Max-Planck Institute for Astrophysics, Karl-Schwarzschild-Straße 1, D-85748 Garching, Germany\\
$^{9}$Department of Physics and Astronomy and PITT PACC, University of Pittsburgh, Pittsburgh, PA 15260, USA\\
$^{10}$ Kavli Institute for Cosmology, University of Cambridge, Madingley Road, Cambridge, CB3 0HA, UK\\
$^{11}$ Cavendish Laboratory, University of Cambridge, 19 JJ Thomson Avenue, Cambridge, CB3 0HE, UK}

\date{Accepted XXX. Received YYY; in original form ZZZ}

\pubyear{2023}

\begin{document}
\label{firstpage}
\pagerange{\pageref{firstpage}--\pageref{lastpage}}
\maketitle

\begin{abstract}
We report the detection of cold dust in an apparently quiescent massive galaxy ($\log({M_{\star}/M_{\odot}})\approx11$) at $z\sim2$ (G4).
The source is identified as a serendipitous 2 mm continuum source in a deep ALMA observation within the field of Q2343-BX610, a $z=2.21$ massive star-forming disk galaxy.
Available multi-band photometry of G4 suggests redshift of $z\sim2$ and a low specific star-formation rate (sSFR), $\log(SFR/M_{\star}) [yr^{-1}] \approx -10.2$, corresponding to $\approx1.2$ dex below the $z=2$ main sequence (MS).
G4 appears to be a peculiar dust-rich quiescent galaxy for its stellar mass ($\log({M_{\rm dust}/M_{\star}}) = -2.71 \pm 0.26$), with its estimated mass-weighted age ($\sim$ 1-2 Gyr).
We compile $z\gtrsim1$ quiescent galaxies in the literature and discuss their age-$\Delta$MS and $\log({M_{\rm dust}/M_{\star}})$-age relations to investigate passive evolution and dust depletion scale.
A long dust depletion time and its morphology suggest morphological quenching along with less efficient feedback that could have acted on G4.
The estimated dust yield for G4 further supports this idea, requiring efficient survival of dust and/or grain growth, and rejuvenation (or additional accretion).
Follow-up observations probing the stellar light and cold dust peak are necessary to understand the implication of these findings in the broader context of galaxy evolutionary studies and quenching in the early universe.
\end{abstract}




\section{Introduction}

The general picture drawn over the last two decades in galaxy evolutionary study is that galaxies show bimodality in their colors, morphologies, and star formation rates (SFRs).
Star-forming galaxies are characterized by bluer colors and disk-like morphology, having rich cold interstellar medium (ISM).
In comparison, quiescent galaxies have redder colors, spheroidal morphology (or higher fraction of bulge component) with less gas (e.g., \citealt{Strateva2001, Kauffmann2003, Baldry2004, Balogh2004, Bell2004, Wuyts2011a, Saintong2011, Young2011}).
In line with this picture, extensive surveys of cold ISM in star-forming galaxies up to $z\sim2$ have shown that the cosmic star-formation activity (e.g., \citealt{Madau2014}) is primarily governed by the available gas within (and outside that falls into) the galaxies (e.g., \citealt{Tacconi2020, Walter2020} and references therein).

Therefore, it has been generally considered that a key aspect of quenching star formation is the reduction of cold ISM reservoirs.
Theories and simulations proposed that it can be achieved by expulsion via outflows or depletion via star formation (e.g., \citealt{Silk1998, DiMatteo2005, Springel2005d}).
Active galactic nuclei (AGN) have been regarded as a preferable agent of outflows in the high mass ($\log(M_{\star}/M_{\odot}) \gtrsim10.7$) regime, which is also observationally indicated at $z\sim1-2$ (e.g., \citealt{ForsterSchreiber2018} and references therein; \citealt{Concas2022}).
Supernova feedback (outflow and shock) would be more efficient in sweeping the gas out of the galaxy in a low mass regime (e.g., \citealt{Dekel1986, Efstathiou2000}).
These two mechanisms were taken to explain the mismatch between the observed stellar mass function and halo mass function high-mass (AGN) and low-mass (supernovae) regimes early on in galaxy evolution studies (e.g., \citealt{White1978, White1991, Balogh2001, Springel2003}; see also reviews of \citealt{Somerville2015, Naab2017}).
A remarkable consistency in the cosmic evolution of the black hole accretion activity further implies a possible role of AGN in the interplay of cosmic gas and star formation evolution.
Additional mechanisms that may contribute to quenching without exhausting the available gas reservoir include heating of the halo gas (via virial shock or kinetic mode feedback from AGN) preventing further infall onto the galaxy by counteracting cooling (e.g., \citealt{Rees1977, Blumenthal1984, Birnboim2003, Bower2006, Croton2006, Somerville2008}) or stabilization by a massive central mass component (e.g., \citealt{Martig2009, Ceverino2010}).

However, at $z\gtrsim1$, when the most massive quiescent galaxies start to form (e.g., \citealt{Thomas2005, Thomas2010}), observations provided inconclusive results to explain the connection between the removal of cold ISM and these potential quenching mechanisms.
Stacking analysis studies found that there is a non-negligible amount of cold dust and gas in quiescent galaxies at $z=1-3$ (\citealt{Gobat2017, Magdis2021, Blanquez-Sese2023}), while individual observations on quiescent galaxies did not offer a convergence with cold gas and dust content in the high redshift quiescent galaxies (\citealt{Sargent2015, Hayashi2018, Belli2021, Whitaker2021, Caliendo2021, Suzuki2022, Kalita2021, Williams2021, Morishita2022, Gobat2022}).
The detection and non-detection of cold gas in post-starburst galaxies at $z>0.5$ also raised a similar question germane to quenching at high redshift (\citealt{Suess2017, Spilker2018, Woodrum2022, Bezanson2022}).
Overall, there is a large scatter of the observed dust and gas content in the individual quiescent and post-starburst galaxies at high redshift.
These observational results perhaps indicate that quenching star formation may not occur from a single mechanism or cannot be explained with simplified bimodal origins, making the full picture of galaxy evolution more complicated.

Whatever the feedback mechanisms are, recent studies advocated the need for bimodal quenching paths in the time scales -- slow and fast -- both in observations (e.g., \citealt{Barro2013, Barro2016a, Schawinski2014, Yesuf2014, Wu2018b, Belli2019, Carnall2019, Tacchella2022}) and simulations (e.g., \citealt{Rodriguez-Montero2019}).
At $z\gtrsim2$, spectroscopically confirmed massive, quiescent galaxies corroborated the fast quenching mode (e.g., \citealt{Glazebrook2017, Forrest2020, Stockmann2020, Valentino2020a}).
How do dust and cold ISM respond to this bimodal quenching time scale and especially fast quenching paths at high redshift? 
And what can we learn from the constraints of observed dust in high redshift quiescent galaxies to understand galaxy evolution?

This work builds upon a very deep observation with ALMA ($\sim20$ hr in total) -- one of the deepest images ever made with ALMA at this redshift -- targeting a main-sequence galaxy at $z=2.21$ (Q2343-BX610).
We detect six continuum sources above peak signal-to-noise ratio ($S/N_{\rm peak}$) of four, five of which were made \textit{serendipitously}.
This Paper delivers the first analysis of these sources, focusing on the discovery of a candidate dust-rich quiescent galaxy at $z\sim2$. 
Comparing its properties with other high redshift quiescent galaxies available, we allude to ideas to answer the questions above.

The Paper is structured as follows. Section~\ref{sec:obs} describes the ALMA observation, data analysis, and flux measurement of the continuum sources. 
Section~\ref{sec:sed} illustrates the multi-wavelength (optical and near-infrared, opt/NIR) data (\S\ref{sec:counterpart}), and SED analysis of the quiescent galaxy candidate.
It includes the photometric redshift estimate (\S\ref{sec:photoz}), and spectral energy distribution (SED) fitting (\S\ref{sec:sed_model}). 
Section~\ref{sec:result} describes overall SED fitting results, SFR constraints (\S\ref{sec:sedsfr}), UVJ color and stellar age (\S\ref{sec:uvj}), dust mass measurements (\S\ref{sec:dts}). 
Section~\ref{sec:discussion} delves into the details of the identified quiescent galaxy, by comparing it with respect to high redshift quiescent galaxies in terms of age and offset from the main-sequence (\S\ref{sec:age-dms}) and dust-to-stellar mass ratio at given age. (\S\ref{sec:dustrichness}). We present a discussion of the potential quenching mechanisms (\S\ref{sec:Qpath}) of G4, followed by the potential origin of the observed dust at $z\gtrsim1$ (\S\ref{sec:dustyield}).
Finally, we conclude and summarize our results in Section~\ref{sec:conclusion}.
We adopt a standard $\Lambda$CDM cosmology with $H_0 =70$ km s$^{-1}$ Mpc$^{-1}$ and $\Omega_m=0.3$ and Chabrier initial mass function (IMF; \citealt{Chabrier2003}).

\section{Observations and source detection}\label{sec:obs}
\begin{figure*}
\centering
\includegraphics[width=0.85\textwidth, bb=0 0 1900 1000]{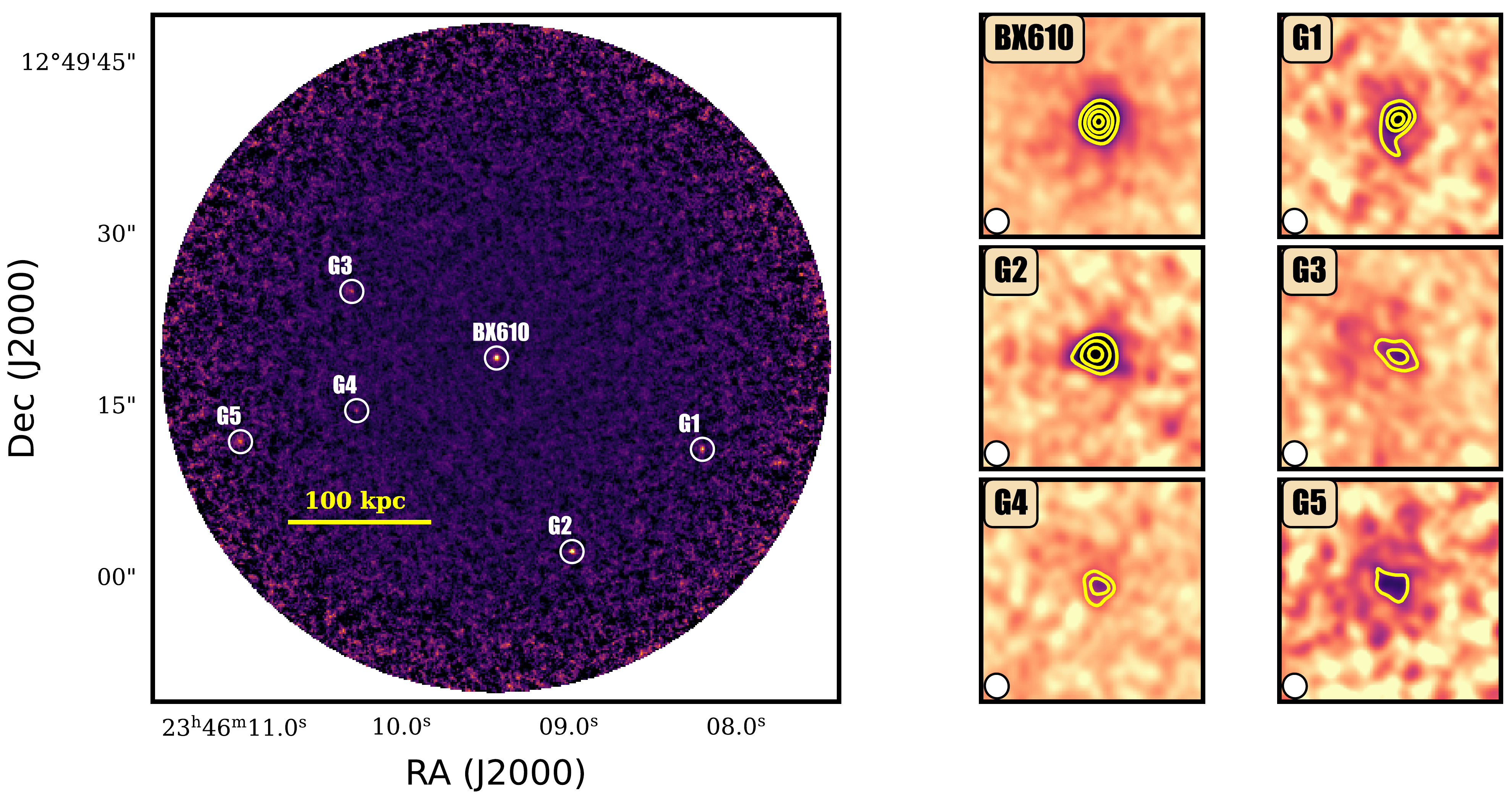}
`\caption{The ALMA Band 4 continuum image near BX610 where we detect six continuum sources at 2~mm. In the left panel, we show the scale bar to indicate 100 kpc (physical) at $z=2.2$, which corresponds to the redshift of BX610 and G1. The right panel size for each galaxy is 2\as$\times$2\as. The contours are shown from 3$\sigma$ in steps of 2$\sigma$, and the beam (0.\as22$\times$0.\as21) is shown on the bottom left for the natural weighting \textsc{CLEAN} map. -3$\sigma$ contours would be shown as a dashed line, which we do not find around the sources.\label{fig:continuum}}
\end{figure*}

The original program aimed to study one of the well-studied, typical main-sequence galaxies at $z=2.21$, Q2343-BX610 (hereafter BX610), at an angular resolution of 0\as.2 (ID: ADS/JAO.ALMA\#2019.1.01362.S, PI: R. Herrera-Camus).
Band~4 receivers were used to cover both \cofour\, and \cione \,(hereafter, \cone)\, emissions to study the cold interstellar medium properties and gas kinematics to resolve the inner structure of the galaxy down to $\sim1.5$ kpc scale, close to the Toomre length at this redshift.
Therefore, antenna configuration was chosen to achieve $\approx$0\as.2 resolution, with baseline lengths ranging between 15 m and 3.7 km using 44-48 antennas.
On-source time was 13.6 hrs ($\simeq$20 hrs in total including the overheads), which enables the characterization of the cold molecular gas kinematics including radial gas inflow in addition to the global ordered rotational motions for the target galaxy BX610.
The detailed analyses on the main target, BX610, are presented by \citet[]{Genzel2023} and S. Arriagada and R. Herrera-Camus et al. 2023 (in preparation). 
This deep integration then allowed us to detect other sources within the ALMA field of view (FoV) accordingly.

Four spectral windows (SPWs) are used, two of each placed in the upper and the lower sideband, respectively, to cover the lines and the underlying continuum.
Two SPWs are set in the Frequency-Division Mode (FDM) to detect the redshifted \cofour\, and \cone\, with a channel width of 3.906MHz ($\sim$ 8.2 km s$^{-1}$) covering 1.875 GHz bandwidth.
The remaining two SPWs are observed in the Time-Division Mode (TDM) with a 2.0-GHz bandwidth at the 15.6-MHz resolution to cover the dust continuum at 2 mm.

We used \textsc{CASA} (\citealt{McMullin2007}) version 6.1.1.15 for the calibration of visibility data and imaging.
For visibility calibration, we used the pipeline script provided by the ALMA Regional Center staff.
Continuum and cube images are produced by \textsc{CASA} task, $\mathtt{tclean}$, and deconvolved down to 2$\sigma$ noise level, which is initially measured from the dirty map.

The continuum image is obtained using the line-free spectral windows observed in the TDM mode.
The synthesized beam is 0\as.23\x0\as.22, 0\as.22\x0\as.20 and  0\as.22\x0\as.21 for \cofour, \cone\, and 2 mm continuum with natural weighting, respectively.
Tapered images are also created by setting \texttt{uvtaper} parameters of 0\as.15 and 0\as.35 in the $\mathtt{tclean}$ task to check the presence of extended emissions. 
We then circularize the beam for these tapered maps, by specifying the \texttt{restoringbeam} parameter as 0\as.29 and 0\as.48 respectively; so the final beams for the tapered images are 0\as.29\x0\as.29 and 0\as.48\x0\as.48.
With the natural weighting, the typical continuum noise level around the phase center (BX610) is $\approx$ 3 $\mu$Jy beam$^{-1}$ and it goes up to $\approx 7$ $\mu$Jy beam$^{-1}$ around the area close to the farthest continuum detected source from the center.

We identified six dust continuum sources with a signal-to-noise ratio of the peak flux greater than four in the ALMA map as shown in Figure~\ref{fig:continuum}.
The ALMA photometry is performed based on a fixed aperture and cross-checked by the result with the 2D Gaussian fitting using \texttt{imfit} and the growth curve.
We took the aperture size of 1\as.0, encompassing most of the emissions from the sources, which are compact.
We list the continuum fluxes for the continuum detected sources in Table~\ref{tab:almaflux}.

We checked the publicly available data on the ALMA archive observed at shallower depths and different configurations (\#2013.1.00059S, 2017.1.00856.S, 2017.1.01045.S) in Band 4. 
We confirmed that BX610's flux values in those programs are consistent with those reported here within the errors\footnote{We note that \#2013.1.00059.S gives $\sim$1.5 times higher value than listed in Table~\ref{tab:almaflux}, but the flux calibrator for this observation was Ceres which had a higher uncertainty in the flux scale (up to $\sim20\%$) according to the QA2 report. \#2017.1.01045.S was obtained at a higher angular resolution (0\as.09\x0\as.07) with $\approx$40\% of missing flux with respect to values reported here.
Our measurement is consistent with a coarser resolution (2\as.3$\times$1\as.86) data in \#2017.1.00856.S which obtained a better flux calibration.}.
We note that the BX610's 2~mm continuum flux measured in \citet{Brisbin2019} is a factor $\sim2$ larger than our measurement at the same frequency.
Although it is unclear whether this difference can be attributed to calibration errors in the flux in the Plateau de Bure Interferometer (PdBI) observations, our flux estimate is the conservative estimate of the total flux.
This ensures our later discussion on the dust mass and dust-to-stellar mass ratio for G4.

We also checked the ALMA archive if there is additional data available at higher frequencies, finding that two ALMA project (\#2015.1.00250.S and \#2019.1.00853.S) covers three galaxies (BX610, G3, and G4) at Band 6 (1.2 mm) within the field of view. 
G3 and G4 are on the edge of the coverage where the sensitivity is degraded by 50-60\% with respect to the central position.
We find $\approx 5\sigma$ and $\approx 3\sigma$ peak features in G3 and G4, respectively using \#2015.1.00250.S.
The resolution of \#2019.1.00853.S's Band 6 program is 0\as.087\x0.\as076 for natural weighting, where we do not find a noticeable signal at the position of G4 but G3.
Photometry of Band 6 is also listed in Table~\ref{tab:almaflux} using the same method applied in the 2 mm map. 
We note, however, that the uncertainty of the photometry is larger due to the low signal-to-noise ratios compared to the 2 mm detection which needs deeper follow-up observations.

This paper will focus on the analyses of G4. Counterpart identification is performed on all objects described in the next section.
We defer the full discussions on other continuum and line-detected sources from the ALMA program to another paper (M. Lee et al. in preparation).
Instead, we present a summary of the spectroscopic redshift of the 2~mm-continuum detected galaxies as follows and in Table~\ref{tab:almaflux}.
Among six ALMA continuum sources, CO and \cone, lines are detected in BX610 and G1, and one line for another (G5).
The redshift of G1 is $2.21$ based on \cofour\ and \cone\ line detections, confirming it being at the same redshift as BX610.
With a single line and the photometric redshift described below, G5 is less likely associated with the BX610-G1 pair system and is rather a background source at $z\approx3.01$.

\begin{table*}
\caption{A summary of ALMA 2~mm detected sources\label{tab:almaflux}}
\begin{tabular}{cccllccccc}
\hline\hline
Source & RA & Dec & $z_{\rm phot}$ &  $z_{\rm spec}$  &  $S_{\rm 2~mm, aperture}^{a}$ & $S_{\rm 2~mm, imfit}^{a}$ & $S_{\rm 1.2~mm, aperture}^{a,b}$ & $S_{\rm 1.2~mm, imfit}^{a,b}$\\ 
 & (deg) & (deg) & 	&	& (mJy)	 & (mJy) \\ \hline
Q2343-BX610	&	356.53932 & 12.82202 & 2.05	& $2.2107 \pm 0.0001$ & $0.234\pm 0.010$		&	$0.205\pm 0.009$ & $1.72 \pm 0.06$ & $1.48\pm 0.04$\\
G1	&	356.53419 & 12.81979 & 2.09 	&  $2.2093 \pm 0.0001^{c}$ & $0.119 \pm0.020$		&	$0.125\pm 0.021$ & -- & -- \\
G2	&	356.53744 & 12.81731 &	--	& --				& $0.173 \pm0.018$		&	$0.166 \pm 0.017$ & -- & --\\
G3	&	356.54292 & 12.82363 & 	--	& --				& $0.077 \pm0.014$		&	$0.072 \pm 0.016$ & $0.50 \pm0.11$ & $0.50\pm 0.15$\\
G4	&	356.5428	&12.82073 &	$2.13^{d}$ &	--				& $0.037 \pm0.013$		&	$0.040\pm 0.009$ & $0.13 \pm0.10$ &  $0.17\pm 0.09$\\
G5	& 	356.54569 & 12.81998 & 0.17	& 3.01?				& $0.141 \pm 0.029$	&	$0.175 \pm 0.061$ & -- & --\\ \hline\hline
\end{tabular}
    \begin{tablenotes}
      \small
      \item[a]a: The continuum flux is measured using the aperture of 1\as.00 in the natural weighting map either based on the aperture photometry and 2D Gaussian fitting using \texttt{imfit} function in \texttt{CASA}. 
      \item[b]b: For 1.2 mm, we used the ALMA archival data (project code: 2015.1.00250.S) with 0\as.32 \x0\as.29 resolution for G4 and G3. For Q2343-BX610, we used a tapered map at 0\as.40 \x0\as.37 based on the investigation of the growth curve.
      \item[c]c: G1's redshift is based on \cofour and \cone, detections and more details of line emitting sources will be presented in M. Lee et al. (in preparation).
      \item[d]d: The $z=2.21$ solution is not rejected. See Section~\ref{sec:photoz} and Figure~\ref{fig:photoz}.
      \end{tablenotes}
\end{table*}%

\section{Photometric redshifts and SED fitting}\label{sec:sed}
\begin{figure}
\centering
\includegraphics[width=0.48\textwidth, bb=0 0 630 760]{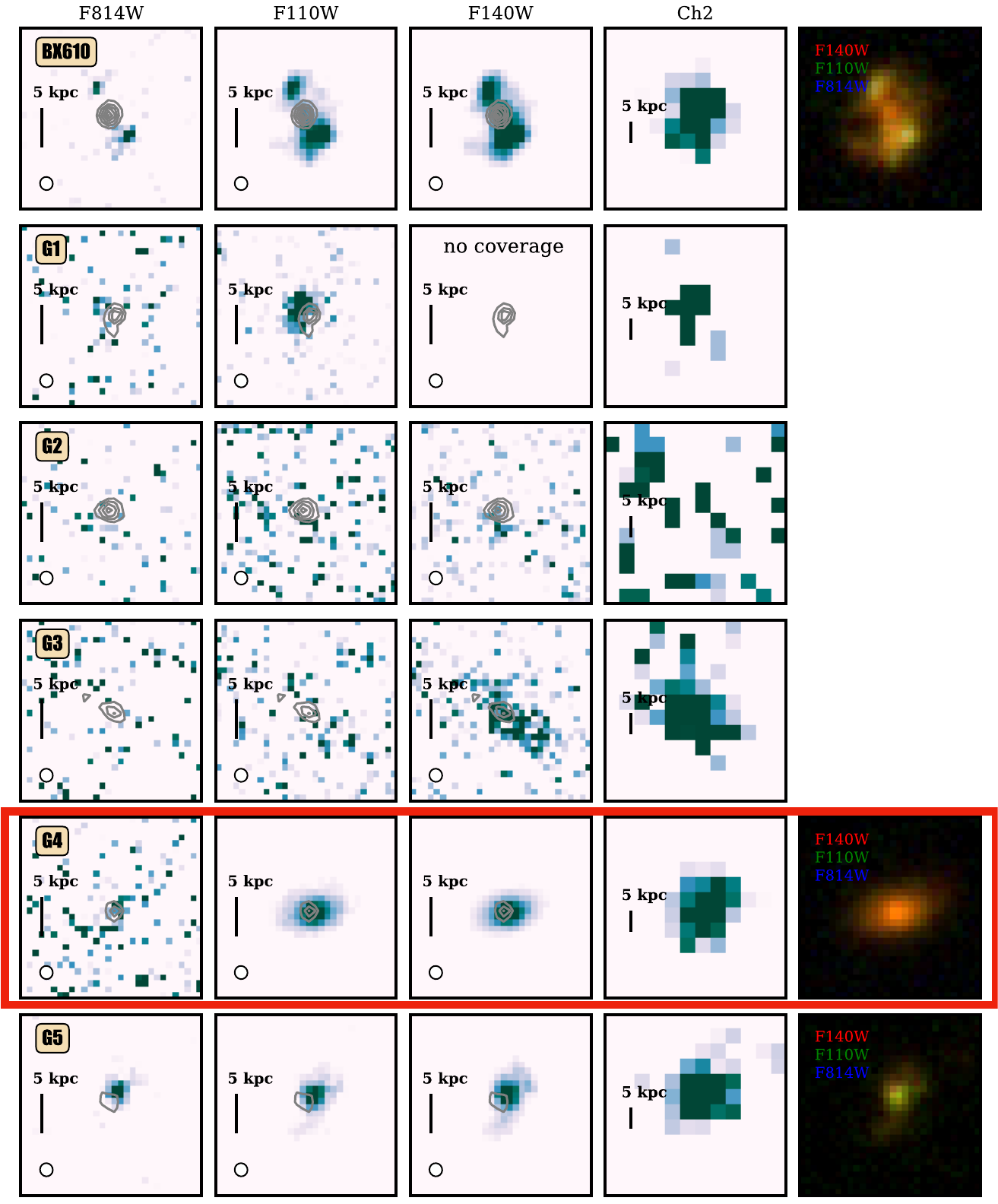}
\caption{Cutout images of \hst/F814W, \hst/F110W, \hst/F140W and \spitzer/IRAC/ch2 and false color images (red: F140W, green: F110W, blue: F814W) for all galaxies detected in ALMA 2 mm continuum. The centers of individual images are all at the 2 mm peaks. The \spitzer\, image is zoomed out to afford the larger pixel size. A scale bar indicating 5 kpc at $z=2.21$ is shown. Overlaid contours are ALMA 2 mm continuum detection (starting from 3$\sigma$ in steps of 2$\sigma$) and the bottom left shows the beam size of the ALMA image (0\as.22\x0\as.21). G4 is highlighted in a red box.\label{fig:hst}}
\end{figure}

\subsection{Counterpart identification and photometry at shorter wavelengths}\label{sec:counterpart}
We gathered all available data from optical and near-/mid- infrared (NIR/MIR) observations from the ground ($U_n$, $G$, \Rs, $J$, $H$, $K_s$) and space (\hst; F814W/F110W/F140W, \spitzer; IRAC Channel 2, MIPS 24\um) and archival data from ALMA.
Due to the very different point-spread functions (PSFs, ranging from $\approx$0\as.15 to $\approx$5\as.5) in different wavelengths, we choose to use different, reasonably large aperture sizes and correct it to recover the total fluxes (so that color corrections have less effect on the photometry for the same aperture correction) described as follows.

$U_n$, $G$, and \Rs\, band images were obtained by the Palomar telescope using the Large Format Camera (LFC) in 2001. 
The description of the data and analysis of the observations (Q2343 field) is available in \citet{Steidel2004} and the references therein. 
The $K_s$ images are also from the Palomar 5m telescope, taken with the Wide Field Infrared Camera (WIRC) in 2003 and 2004. 
We refer the readers to \citet{Erb2006} for the details of the observations and data analyses.
We used $J$ and $H$ band images taken from Magellan/FourStar, obtained in 2013. 
The $J$ image is a stack of the $J2$ and $J3$ intermediate bands, and the $H$ band is a stack of $H1$ and $H2$.  
All of the Magellan data were taken by Gwen Rudie.
All of the above photometry was performed using \textsc{SExtractor}, after matching the point-spread function (PSF) in all bands (PSF $\sim 1$\as) using the $R_s$ (for most objects) or $H$ band (for very red objects) image as the detection band in ``dual-band" mode. 
This strategy is to optimize the source detection and the corresponding isophotal aperture given that UV color-selected galaxies, the original main target of the observations, are better detected in the optical band than in the NIR, while the red objects, like G4, are conversely detected in the NIR bands with higher S/N.
The magnitudes were corrected from isophotal to ``total" using the \textsc{SExtractor} \texttt{mag$\_$auto} in the relevant detection band.

For \hst, we obtained the imaging data from the MAST archive processed with \textsc{Grizli}\footnote{\url{https://github.com/gbrammer/grizli}} pipeline \citep{Brammer2021, Brammer2022a}.
The original work using the WFC3/F110W image was presented in \citet{Tacchella2015b}.
We used the \spitzer\, data from the \textsc{golfir}\footnote{\url{https://github.com/gbrammer/golfir}} project \citep{Brammer2022b} which generated the \spitzer\, mosaics from the pipeline processed data.
The maps were corrected for the astrometry.

For the HST bands and IRAC 4.5$\mu$m photometry, we used \textsc{SEP}\footnote{\url{https://github.com/kbarbary/sep}, SEP is a fork of original \textsc{SExtractor} and uses the same core algorithms of \textsc{SExtractor}.} (version 1.2.0) to extract sources.
For the same reason of different PSF, the source extraction and photometry measurement were done independently between the \hst\, and \spitzer\, maps.
For the \hst\, bands, we used the longest wavelength (F110W (G1) or F140W (other galaxies)) available for source detection.
We used an aperture of 0\as.7 for the photometry in all bands and corrected it to ``total" values within an elliptical Kron aperture \citep{Kron1980}.
The latter was calculated based on the longest \hst\, bands (F140W or F110W), and the same amount of aperture correction is applied for all the other bands. 
Given the aperture size is large enough the color correction is negligible.
For the IRAC photometry, we used a larger aperture of 3\as.6 and applied the aperture correction which ranges between 10-18\%.
We also checked the curve of growth of the detected galaxies in the IRAC map, confirming that the aperture correction and thus the estimated flux is reasonable.

The MIPS 24$\mu$m data was taken in 2006 and the description of the data reduction is available in \citet{Reddy2010}. 
The MIPS photometry was done using an aperture of 3\as centered at the 2 mm sources, and an aperture correction factor of 3.097 was applied based on the MIPS Instrument Handbook (v3.2.1) Table.4.13. 
We obtained the errors based on the median value of standard deviation values given the same aperture taken from random, emission-free positions around the source.

Counterpart identification was performed based on the position of the 2 mm emission.
We identified the optical/NIR/MIR counterpart at 0\as.10 - 0\as.24 offset from the 2 mm peak positions except for G2.
Considering the beam and the point-spread function (PSF) sizes of ALMA and optical/NIR observations, and the signal-to-noise ratio, we regard the offset as negligible to pin down the counterparts.
The offsets would rather be originated from the internal structures and different levels of dust extinction across the galaxy.
Finally, we note that there is no additional counterpart identified in the HST imaging within the IRAC and MIPS apertures that could contribute to the photometry.

Figure~\ref{fig:hst} shows the three \hst\ filter and \spitzer/IRAC Ch2 (4.5$\mu$m) maps overlaying the 2 mm ALMA continuum contours.
G2 does not have clear counterparts in any of the bands which may be partly owing to the poor sensitivity (i.e., on the edge of the coverage) around the source. 
G3 starts to appear in the J, F140W maps and the longer wavelengths, which may imply either an obscured (indicated by the dust detection) or a Lyman/Balmer-break galaxy at higher redshift.
G2 and G3 have insufficient data points to perform the full SED fitting and get a good photometric redshift estimate. 
Follow-up observations of these two are necessary to characterize these two optically dark/faint galaxies.
Hereafter, we consider only G4 for further analyses.

\subsection{Photometric redshift}\label{sec:photoz}

We use \textsc{eazy-py}\footnote{\url{https://github.com/gbrammer/eazy-py}} \citep{Brammer2008} to get a first handle on the photo-$z$ estimate.
Full details concerning \textsc{eazy-py} and the template set can be found in \citet{Brammer2008, Kokorev2022}; see also \citet{Blanton2007} for the template creation algorithm.
Briefly, the best-fit photo-$z$ is estimated from a linear combination of a set of 13 templates from the Flexible Stellar Populations Synthesis code (\textsc{fsps}, \citealt{Conroy2010}) 
and one additional template recently obtained from $z=8.5$ -- ID4590 from \citet{Carnall2022} (namely, \texttt{carnall$\_$sfhz$\_$13} template in \textsc{eazy-py})\footnote{We examined the fitting only with thirteen templates \texttt{corr$\_$sfhz$\_$13} from the FSPS, but we find the inclusion of this template has a negligible impact on the redshift of the galaxies we consider here.}.
The first thirteen templates are constructed based on redshift-dependent SFHs (a combination of four star-formation histories) and 2-3 dust models.
Star-formation histories that start earlier than the age of the Universe are excluded at a given redshift.
We note that there is a limitation of this set-up for \textsc{eazy-py} that there will be $z$-SFH degeneracies if there are an insufficient number of bands. 
We explore different SFHs using different SED fitting codes in Appendix~\ref{app:sed} which does not alter our conclusion on G4.
The set of thirteen templates is chosen based on an algorithm to reasonably cover the rest-frame color space following the philosophy of \citet{Brammer2008}.
The additional template can better explain the extremely strong emission lines seen in early JWST spectra of z>6 galaxies.
Templates are constructed on the \citet{Chabrier2003} IMF.
\citet{Kriek2013} dust attenuation law (dust index $\delta$ = -0.1, $R_{\rm V} = 3.1$) is adopted and its maximum attenuation is set to be redshift-dependent.
On the \textsc{fsps} thirteen templates, a fixed grid of nebular emission lines and continuum from \textsc{cloudy} (v13.03) models are added (metallicity: $\log{(Z/Z_{\odot})} \in [-1.2, 0]$, ionization parameter $\log{(U)} \in [-1.64, -2]$\footnote{It is a high-value range for $z\sim2$ star-forming galaxies (e.g., \citealt{Steidel2016, Strom2017}), but inclusion of this does not change our fit of G4, where the contribution of nebular emission is negligible.}; \citealt{Byler2018}).
Given the unconstrained nature of fixed parameters used in \textsc{eazy-py}, we will only use the photo-$z$ information from \textsc{eazy-py}.
Finally, a correction for the effect of dust in the Milky Way ($E(B-V) = 0.0327$) is applied to the templates within \textsc{eazy-py}, pulling the Galactic dust map by \citet{Schlafly2011} from \textsc{dustmaps} \citep{Green2018}. 
For the fitting process, we set minimum and maximum redshifts to 0.01 and 12, respectively, and the redshift step (\texttt{z\_step}) to 0.01.
Referring to \citet{Brammer2008}, we use the extended K-band prior (\texttt{prior\_K\_extend.dat}) to lower the weight of the low redshift solution.

\begin{figure*}
\centering
\includegraphics[width=0.80\textwidth, bb=0 0 2500 1300]{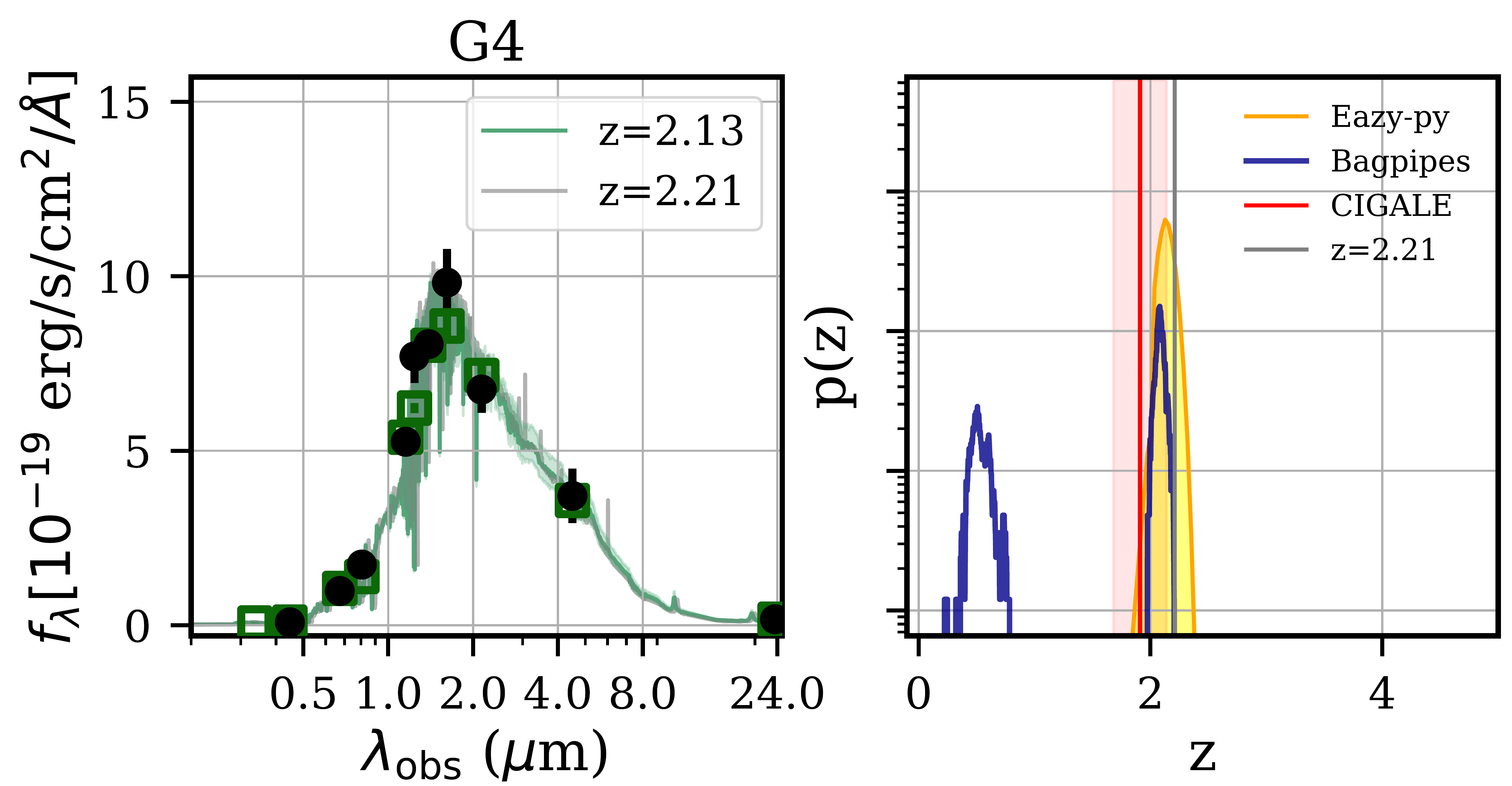}
\caption{Photometric redshift estimation from SED fitting for G4. The best-fit SED template fit from \textsc{eazy-py} is shown on its left panel, and the best-fit redshift is indicated as green lines. The grey line is the best fit at a fixed redshift of $z=2.21$, the spectroscopic redshift of BX610 and G1. Black-filled circles show the observed flux and open squares show the best-fit model convolved with the filter response. 
On the right panel, $P(z)$ distribution of the \textsc{eazy-py} (yellow shaded) and \textsc{bagpieps} (dark blue, uniform redshift prior), and \textsc{cigale} (red shaded) are shown. See also Table~\ref{tab:sedzphot} in Appendix~\ref{app:sed}.\label{fig:photoz}}
\end{figure*}

To gain more credentials of photo-$z$ estimate from \textsc{eazy-py}, we also run \textsc{bagpipes} and \textsc{cigale} using different SFH.
The details of the SED results and the set-up to fit redshift are summarized in Appendix~\ref{app:sed}.
In general, they agree well with each other, giving the best-fit photometric redshift of $z\simeq2$ for G4.
We overlay the final results in Figure~\ref{fig:photoz} for the $P(z)$ distribution from \textsc{eazy-py}, \textsc{bagpipes}, and \textsc{cigale} photo-z constraint.
Hereafter, we consider G4 to be at $z\sim2$.

\subsection{SED fitting}\label{sec:sed_model}
Given the consistency between the estimated photo-$z$ from different SED codes, we use \textsc{bagpipes}\footnote{\url{https://github.com/ACCarnall/bagpipes}} \citep{Carnall2018}, \textsc{fast++}\footnote{\url{https://github.com/cschreib/fastpp}} \citep{Kriek2009, Schreiber2018b} and \textsc{cigale}\footnote{\url{https://cigale.lam.fr/}} \citep{Boquien2019} by fixing the redshift based on the best-fit photometric redshift from \textsc{eazy-py} ($z=2.13$). 
The photometric data used for G4's SED fitting is summarized in Table~\ref{tab:phot_gal4}.
We use three different codes to see if these codes are able to fit the G4's photometry and thus strengthen the reliability of the overall results.
We regard the results with fixed redshift as the best estimate from SED, which is summarized in Table~\ref{tab:sedbestfit}. 
In Appexdix~\ref{app:sed}, we explore the impact of fitting the photometry with the redshift left free to vary. The comparison indicates reasonable agreement between parameters derived with and without fixing the redshift, and does not alter the overall conclusion that G4 is a massive quiescent galaxy at $z\sim2$ with low specific star-formation rate $\log({sSFR/{\rm yr^{-1}}})\lesssim-12.5$. 

\subsubsection{\textsc{bagpipes}}\label{sec:bagpipes}
We take into account two different types of parametric star-formation histories (SFH) to fit the data points: double-power law SFH ($SFR(t)\propto \big[ (t/\tau)^{\alpha} + (t/\tau)^{-\beta}\big]^{-1}$) and non-parametric SFH.
These two sets of SFH were chosen because an exponentially declining SFH ($\tau$-model) model did not deliver a good fit for G4 using \textsc{bagpipes} giving high $\chi^2$ without reasonable convergence.
Although the $\tau$-model (e.g., \citealt{Schmidt1959, Conroy2013}) is the most popular model that has been extensively used in studies based on large surveys (e.g., \citealt{Ilbert2013, Skelton2014, Stefanon2017}) recent studies advocated the need for more complex or different models to explain the star-formation histories of galaxies (e.g., \citealt{Maraston2010, Wuyts2011a, Pacifici2016, Carnall2018, Leja2019a}).

The double-power law SFH has more flexibility with an additional parameter (a total of 6 parameters) compared to a $\tau$-model.
We vary $\alpha$ and $\beta$ values between 0.01 and 10000, with uniform priors in log space (``\texttt{log$\_$10}"). 
The age of the Universe the peak of star formation ($\tau$) is sampled between 0.1 and 13.5 Gyr and the metallicity range is between 0 and 4.0 $Z_{\odot}$.
We adopt the reddening law of \citet{Calzetti2000} with the $V$-band attenuation ($A_V$) between 0 and 4 magnitudes.
Nebular emission is not taken into account for G4; we tested nebular emission with G4 and the nebular emission did not change our analysis.
We allow the stellar mass formed by the assumed SFH in $7.0<\log({M_{\star}/M_{\odot}})<13.5$.

We also test with non-parametric SFH with continuity prior using \textsc{bagpipes} to test if non-parametric SFH can produce better results.
By its construction, the non-parametric continuity SFH priors can yield a better fit and may overfit because we have more free parameters to fit.
To set the SFH bin, we follow the methodology described in \citet{Leja2019a} with six bins.
The first two recent bins are set to 0-30, and 30-100 Myr, and the remaining four bins are equally spaced in logarithmic time.
\citet{Leja2019a} showed that the number of bins in the mock analysis is insensitive as long as $N_{\rm bins}>4$.
We also confirm that the number of bins has a negligible impact on stellar mass, age, and $
A_{\rm V}$, and SFR for G4 for $4<N_{\rm bins}<9$.
Therefore, we adopt the bin number $N_{\rm bins}=6$, which is computationally less expensive without excessively overfitting given the number of data points.
For G4, we get the consistent stellar mass, and SFR, placing a galaxy far below the star-forming main sequence at $z=2$, as summarized in Table~\ref{tab:sedbestfit}.

\subsubsection{\textsc{fast++}}\label{sec:fast++}
We run \textsc{fast++} with the following assumptions: \citet{Bruzual2003} library (\texttt{bc03}) assuming Chabrier (\citealt{Chabrier2003}) initial mass function (IMF). 
We assume a delayed-$\tau$ model for the sake of simplicity and to allow more flexibility than the $\tau$ model.
The minimum $e$-folding time is set to $\log\tau_{\rm min} = 8.0$ (100 Myr) and maximum of $\log\tau_{\rm max} =10.1$ (12.5 Gyr) in steps of $\log\tau = 0.1$
The minimum time since the onset of star formation is 50 Myr and the maximum is the age of the universe. 
The $\log$(age) grid is made in steps of 0.05.
Solar metallicity and the Calzetti dust attenuation law (\citealt{Calzetti2000}) are adopted with visual extinctions in the range $0<A_V<4$.

\subsubsection{\textsc{cigale}}\label{sec:cigale}
\begin{figure}
\centering
\includegraphics[width=0.48\textwidth, bb=0 0 600 520]{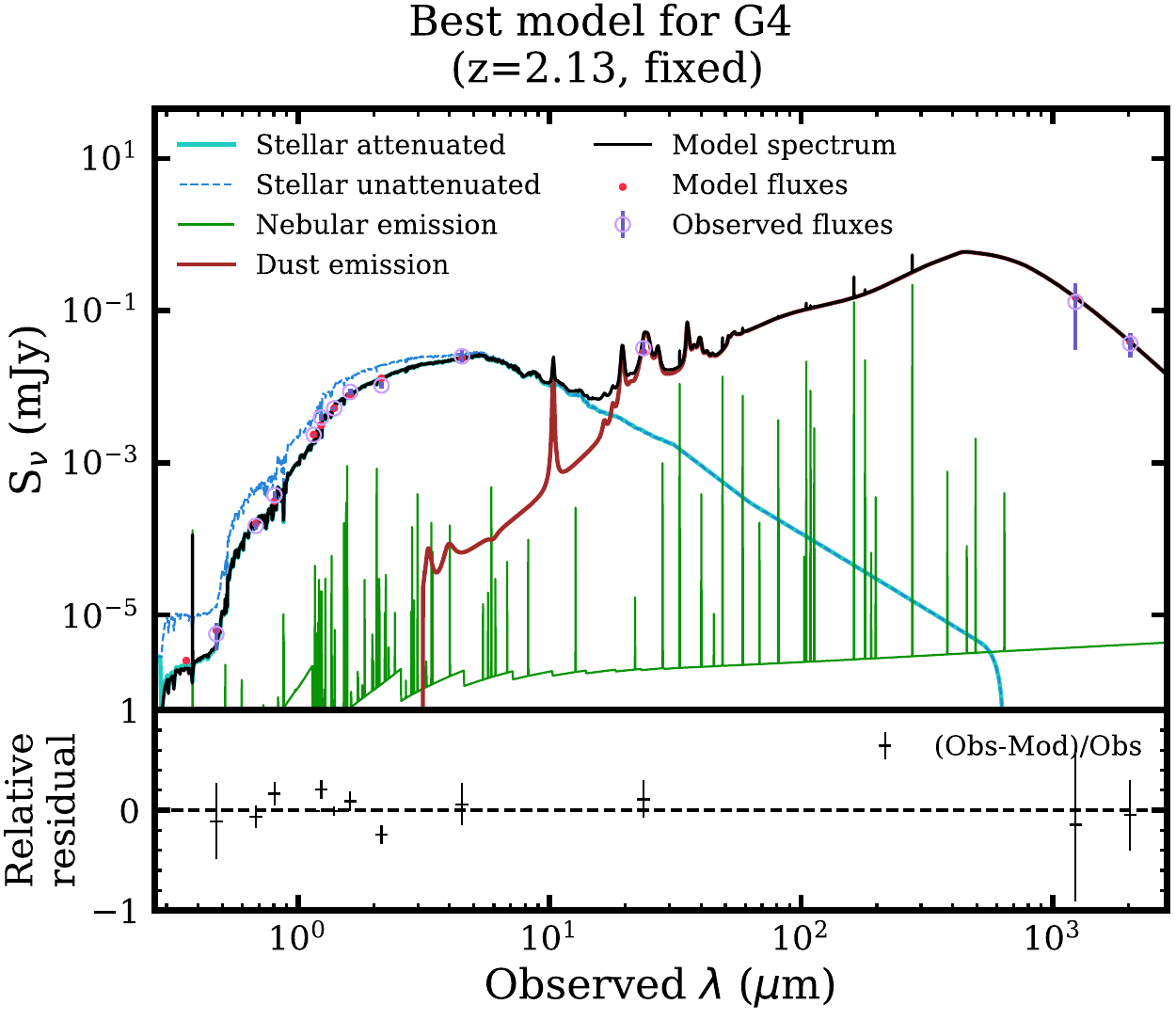}
\caption{\textsc{cigale} best fit SED for G4.\label{fig:cigale_gal4}}
\end{figure}

We use \textsc{cigale} (\citealt{Boquien2019}) in order to take into account the ALMA data points (2 mm and 1.2 mm) to avoid the potential underestimation of the SFR from optical to mid-IR photometry alone, especially for an obscured star-forming case.
We adopt the delayed-$\tau$ model for the SFH.
The other set-up is similar to those used in the other SED fitting process that we used \citet{Bruzual2003} library assuming Chabrier (\citealt{Chabrier2003}) initial mass function (IMF) and use the Calzetti dust attenuation law (\citealt{Calzetti2000}).
As for the dust emission, we use the \citet{Draine2014} model constructed by varying the PAH mass fraction ($q_{\rm PAH} = 1.12-3.90$), minimum radiation field ($U_{\rm min} = 0.5-10$), and power law slope ($\alpha = 1.0-2.0$), which describes the distribution of the radiation field per mass.
We also set the metallicity and nebular emission to vary.
\textsc{cigale} fits G4's SED well with reasonable $\chi^{2} \approx 11.3$ (Figure~\ref{fig:cigale_gal4}).\\

\begin{table*}
\begin{center}
\caption{Opt/NIR Photometric data points for G4\label{tab:phot_gal4}}
\begin{tabular}{ccccccc}
Photometric Band & Un & G & Rs & f814w & f110w & J \\ \hline
Flux ($\mu$Jy) & - & 0.006$\pm$0.002 & 0.149$\pm$0.017 & 0.377$\pm$0.044 & 2.333$\pm$0.022 & 3.945$\pm$0.381 \\ \hline\hline
Photometric Band & f140w & H & Ks & IRAC2 & MIPS24  \\ \hline
Flux ($\mu$Jy) & 5.200$\pm$0.245 & 8.551$\pm$0.825 & 10.375$\pm$1.001 & 25.060$\pm$5.250 & 32.000$\pm$7.600 \\
\end{tabular}
\end{center}
\end{table*}

\begin{table*}
\begin{center}
\caption{Best-fit parameters of G4 from SED fitting\label{tab:sedbestfit}}
\begin{tabular}{c|cccc}
\hline \hline
\multicolumn{5}{c}{G4} \\ \hline
& FAST++ & Bagpipes &  Bagpipes & CIGALE \\ \hline
Redshift		&	2.13$^{a}$		& 2.13$^{a}$  &	2.13$^{a}$	&	2.13$^{a}$	\\
$\log M_{\star}$	&	$10.99\pm0.05$	&	11.16		&    11.22   &	$11.25\pm0.10$	\\
				&				& [11.10,11.21]& [11.17,11.30]	 &	\\
SFR$_{\rm SED}^{\dagger}$&		0.02		&	0.00$^{c}$		& 0.01$^{c}$ &	1.3e-4$\pm$7.04e-5$^{c,d}$	\\
			&					& [0.00,0.00]	& [0.00,0.04]	&	\\
log($\Delta$MS)$^{\dagger}$	&		-3.79			&	$-69.68$	& -4.40   &	-6.20 \\
$\log$(age)$^{b}$	&		9.15			&	8.83	& 9.27	& 	$9.26\pm0.20$	\\ 
			&			& [8.79,8.97] & [9.20,9.30]	& \\
SFH		& delayed-$\tau$ & double-power law & non-parametric &  	delayed-$\tau$	\\ 
$A_{\rm V}$ & 0.10 & 0.75 &  0.51    & $0.57\pm 0.10$\\
            &       & [0.50,0.93] & [0.33,0.72] &   \\ 
$\chi^{2\ddagger}$    &    2.1   &   15.8    &  22.4 & 11.3    \\ \hline \hline
\end{tabular}
    \begin{tablenotes}
      \small
     \item[$\dagger$] $^{\dagger}$: We note the limitation of the SFR from the SED which is likely to introduce an unrealistically low value of SFR and hence $\Delta$MS especially for the parametric SFH models. The photometry (\Rs-band) and the best-fit dust attenuation ($A_{\rm V}$) suggest the SFR of $\sim8\, M_{\odot}$ yr$^{-1}$ and $\log(\Delta$MS) $\sim -1.6$ (see also the main text in Section~\ref{sec:sedsfr}).
     \item[$\ddagger$]  $^{\ddagger}$: We quote the absolute $\chi^{2}$ value, as the templates employed (as is the case for many SED fitting codes) are not independent of each other, and degrees of freedom are ill-defined (e.g., \citealt{Smith2012}).
      \item[a] $^{a}$: best-fit photo-$z$ from \textsc{eazy-py}.
      \item[b] $^{b}$: mass-weighted.
      \item[c] $^{c}$: averaged over 100 Myr
      \item[d] $^{d}$: including the ALMA 2~mm data point.

      \end{tablenotes}
                \end{center}
\end{table*}

\begin{table}
\begin{center}
    \caption{SFR estimate from 2 mm and 24 $\mu$m data points\label{tab:sfr}}
\begin{tabular}{ccccc}
\hline \hline
Method & 2 mm + MBB$^{a}$ & Rest-8$\mu$m$^{b}$ & 24$\mu$m scaling$^{c}$\\ \hline
SFR ($M_{\odot}$ yr$^{-1}$) & $\sim$2-136 & $\sim$30 (16-55) & $\sim$8-10 \\ \hline
\end{tabular}
    \begin{tablenotes}
      \small
      \item[a] $^{a}$: Assuming $\beta = 1.5-2.2$, $T_{\rm d}$ = 17-35 K for modified black body and using \citet{Kennicutt1998a} for $L_{\rm IR}$ to SFR conversion. We note that the 1.2 mm data point gives a range of $2-72\, M_{\odot}$ yr$^{-1}$, but the uncertainty is much larger if we take into account the photometry uncertainty.
      \item[b] $^{b}$: Based on \citet{Reddy2010} and the values in the parenthesis are taking the scatter of the relation between $\nu L_{\nu}$ and L(H$\alpha$) (and SFR), not the 24$\mu$m flux errors.
      \item[c] $^{c}$: Based on the 24 $\mu$m flux scaling between BX610 and G4 and using the SFR of BX610 from using \citet{ForsterSchreiber2018} (115 $M_{\odot}$ yr$^{-1}$) and \citet{Brisbin2019} (140 $M_{\odot}$ yr$^{-1}$).
      \end{tablenotes}
      
\end{center}
\end{table}


\section{Results}\label{sec:result}

\subsection{Consistency in the SED fitting results and SFR from other methods}\label{sec:sedsfr}
~~G4 is a massive galaxy with $\log{(M_{\star}/M_{\odot})}\approx11.0$ based on the SED fitting.
Best-fit SED results of G4 are summarized in Table~\ref{tab:sedbestfit}.
The stellar masses between different codes are overall in good agreement.
The agreement in the stellar mass between different models is consistent with earlier studies that stellar masses are the most robust parameter estimated from SED fitting for ``normal" (and not so blue) galaxies (e.g., \citealt{Papovich2001, Shapley2001, Conroy2013, Pacifici2023}).

All codes yield broadly consistent results favoring the quiescent nature of G4 though there are different levels of inactivity from the SED fitting.
The multiple SED codes indicate that G4 is characterized as a quiescent galaxy with a low level of star-formation activity with $\log({sSFR/{\rm yr^{-1}}})\lesssim-12.5$, which is -3.5 dex lower than the star-forming main sequence at $z=2$.
The \textsc{cigale} fit, where we include the ALMA data points for the fitting with energy-balance assumption also gives a low level of star-formation and a quiescent solution for G4 with a reasonable $\chi^{2}$ value.
The fitting results do not change with and without the 1.2 mm data point for G4.
We show the best fit from \textsc{cigale} in Figure~\ref{fig:cigale_gal4} and the other best-fit SED models in Appendix~\ref{app:sed}.

However, the SFR from the SED could introduce artificially low SFR, especially for the parametric SFH models by its construction.
In this regard, the attenuation corrected \Rs-band (i.e., rest-frame 2000 \AA) magnitude suggests the SFR of $\sim 8\, M_{\odot}$ yr$^{-1}$ and $\log({sSFR/yr^{-1}})\sim$-10.3 instead: we applied the implied extinction correction ($A_{\rm V}\sim0.5$) assuming the Calzetti law, i.e., $A$(2000\AA)$\sim 9\times E({\rm B-V}) \sim 1.5$ mag, which gives $R\sim24.5$ mag and converted this into SFR attributing the emission to UV continuum of young stars.

Given the different degrees of low SFR from the SED best fit and inferred SFR from the \Rs-band photometry, we estimate the SFR from three other different ways using 24 $\mu$m and 2 mm fluxes.
These are more reliable tracers of SFR because they are not affected by dust obscuration. 
The estimated dust extinction ($A_{\rm V}$) is moderate, which can be attributed mainly to the old stellar population ($\approx 0.4$ in $A_{\rm V}$), but it is still good to check whether other different SFR measurements agree with each other.
A summary of underlying estimates is presented in Table~\ref{tab:sfr}

If we use the modified black body (MBB) and take the 2 mm flux, the obscured star formation can range $\approx2-136 \, M_{\odot}$ yr$^{-1}$ assuming $T_{\rm d}$ = 17-35 K and $\beta$ = 1.5-2.2, using \citet{Kennicutt1998a} to convert $L_{\rm IR}$ to SFR and dividing it by 1.6 for Chabrier IMF (\citealt{Madau2014}). 
Here, we assume no contribution from old stellar populations heating the dust.
In general, however, there is a non-negligible contribution of the old stellar population in dust emission for quiescent galaxies, which could play another source of uncertainty constraining the SFR of G4 from $L_{\rm IR}$ and in effect the SFR can be lower than this.
Further, without constraining higher frequency near the dust SED peak and the dust temperature, there is considerable uncertainty in inferring the actual star-formation rate (Table~\ref{tab:sfr}).
ALMA Band 4 covers the rest-frame 660$\mu$m at $z\sim2$, and should mainly trace the cold dust component rather than the SFR. 
Low S/N ($\approx1.5\sigma$ for total flux, Table~\ref{tab:almaflux}) of the 1.2 mm data point does not provide a strong indication of extremely high star formation, giving a similar range of $2-72\, M_{\odot}$ yr$^{-1}$. 
Future high-frequency observations at ALMA Band 9, and 10 tracing the dust peak will help to pin down the SFR.

\citet{Reddy2010} used the MIPS 24$\mu$m flux for $\langle z\rangle\sim2$ galaxies to constrain the relationship between the rest-frame 8$\mu$m luminosity, which traces the emission from polycyclic aromatic hydrocarbons (PAHs), and SFR.
Using our 24$\mu$m flux in Table~\ref{tab:phot_gal4} and equation (2) in \citet{Reddy2010}, we get SFR of $\sim 30 \,M_{\odot}\,{\rm yr}^{-1}$ (total dust-corrected SFR).
The scatter of this fit is about 0.24 dex, giving a range of 16-55 $M_{\odot}$ yr$^{-1}$ from the 24$\mu$m flux.
However, the relationship between rest-frame 8$\mu$m and SFR is constructed based on galaxies forming stars at a higher rate, which might be a source of uncertainty playing a role in this conversion.

The third approach is to estimate SFR by scaling the 24$\mu$m flux with respect to Q2343-BX610, which has a better constraint with more photometric data points from UV-to-FIR, and use the latest SFR measurements of Q2343-BX610.
\citet{Brisbin2019} measured the star-formation rate of BX610 to be 140 $M_{\odot}$ yr$^{-1}$ from their SED fitting including a few FIR data points.
They estimated slightly higher than attenuation corrected SFR from UV-to-NIR (60 $M_{\odot}$ yr$^{-1}$) and H$\alpha$ (115 $M_{\odot}$ yr$^{-1}$) (\citealt{ForsterSchreiber2014, ForsterSchreiber2018}). 
By taking the MIPS 24$\mu$m photometry for BX610 and G4, we estimate the G4's SFR is $\sim$ 8-10 $M_{\odot}$ yr$^{-1}$. 

SFR from different tracers can change by two orders of magnitude (or more; \citealt{Schreiber2018b, Belli2019, Man2021}) between $SFR_{\rm SED}$, $SFR_{\rm IR+UV}$, and $SFR_{\rm [OII]\, or\, H\beta}$ for high redshift quiescent galaxies.
Based on multiple approaches to estimate the SFR, we suggest that the SFR of G4 from the scaling from the 24$\mu$m point (the third method, SFR$\sim$10 $M_{\odot}$ yr$^{-1}$) as the most reasonable (or robust) estimate with available data.
Inferring SFR with only a single data point from 2~mm needs heavy interpolation and the conversion between rest-8$\mu$m to SFR might have unknown systematics for low luminosity regime.
This also minimizes our concern of potential underestimation of SFR from the SED fitting.
This also agrees with the dust attenuation corrected SFR from the rest-frame UV magnitude.
Finally, the adopted SFR of 10 $M_{\odot}$ yr$^{-1}$ also places the galaxy to be in (somewhat) better agreement with the observations and simulations in the age-$\Delta$MS plane (Section~\ref{sec:age-dms}).

With the SFR value of $\sim10$ $M_{\odot}$ yr$^{-1}$, G4 is a quiescent galaxy with lower sSFR ($\log{(sSFR/{\rm yr}^{-1})} \approx -10.2$) at least by $\approx1.2$ dex from the $z=2$ main-sequence, compared to main-sequence galaxies at $z\sim2$.
We note that $\log{(sSFR/{\rm yr}^{-1})} = -10.0\,(-10.5)$ at $z\approx2$ corresponds to the mass-doubling time being more than 3 (10) times the age of the universe at that redshift.
One of the definitions of quiescent galaxies used in the literature is sSFR smaller than $0.3/t_{\rm Hubble}$ \citep{Franx2008}, and G4 satisfies this.
We use this sSFR ($\log{(sSFR/{\rm yr}^{-1})} = -10.2$) as a proxy of the galaxy's quiescence.

We note that available data can not completely reject the possibility of a higher star-formation rate than currently best guessed.
Future observations at higher frequencies at submm covering the dust SED peak are needed to highlight potential obscured star formation that we may still miss.

As for the other parameters, we take the \textsc{bagpipes} results using non-parametric SFH in the following discussions, because the results are close to the median values of four different SED fitting results.
Also, the SFR of G4 from \textsc{bagpipes} non-parametric gives one of the highest star-formation rates that we use as the upper limit constraint of the SFR (from the SED).

\subsection{UVJ color and stellar age}\label{sec:uvj}

\begin{figure}
\centering
\includegraphics[width=0.48\textwidth, bb=0 0 1200 950]{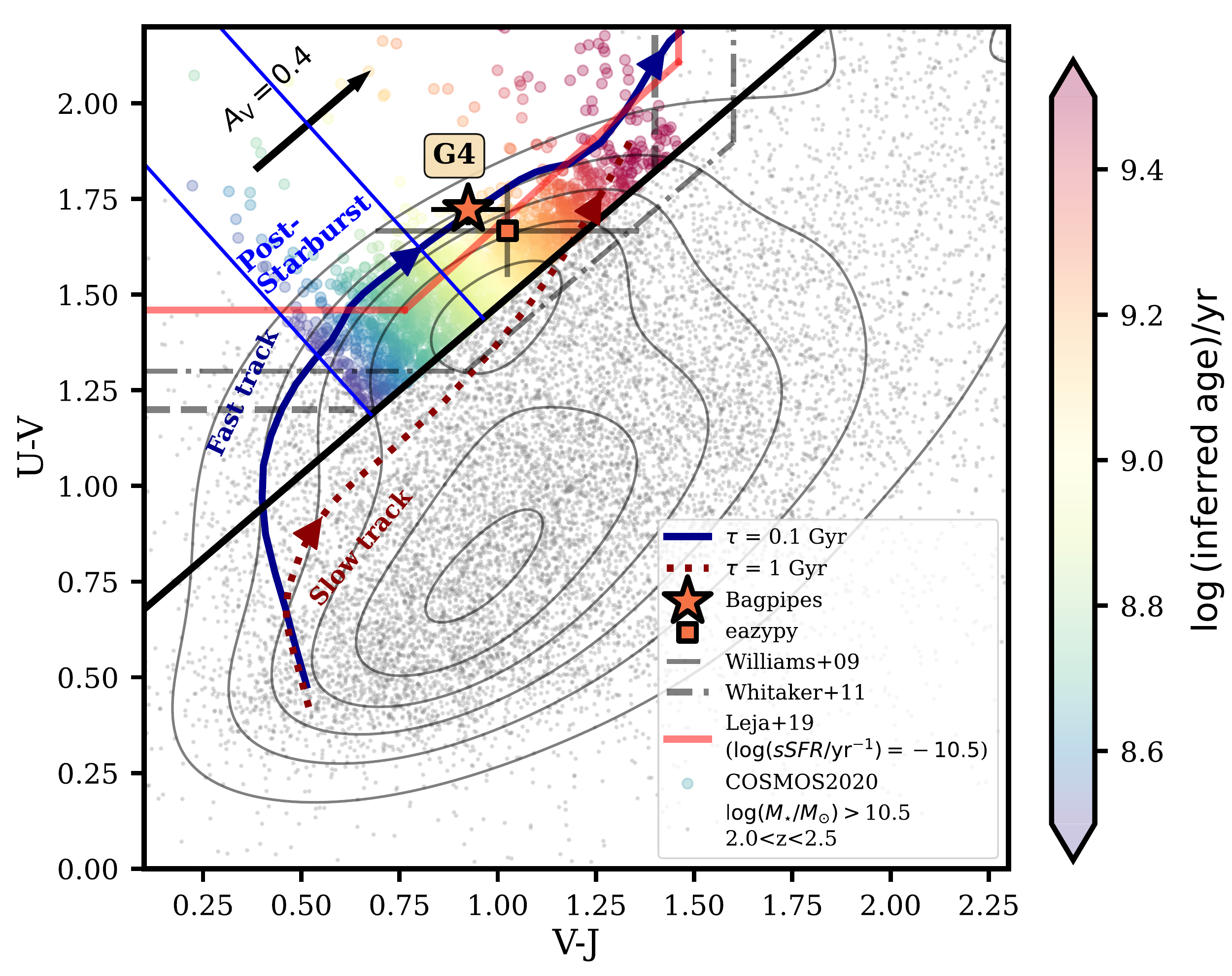}
\caption{The rest-frame $UVJ$ colors of G4 based on the \textsc{bagpipes} (non-parametric SFH) best-fit (star symbol), color-coded by the mass-weighted age obtained from the SED fitting.
The overlaying data points are $\log(M_{\star}/M_{\odot})>10.5$ galaxies at $2.0<z<2.5$ in the COSMOS2020 catalog \citep{Weaver2020} and their number density distributions in contours. 
\textsc{eazy-py} best-fit $UVJ$ colors of G4 are shown as a square for a fair comparison with COSMOS2020.
We define a quiescent region based on \citet{Whitaker2011}.
A thick solid diagonal line represents the definition by \citet{Whitaker2011} with additional cut in the $U-V$ and $V-J$ colors shown as a dashed line.  
COSOMOS2020 galaxies in the quiescent region are color-coded with their expected stellar age based on the empirical age-color relation derived by \citet{Belli2019} (for age $>300$ Myr). 
This shows a good match between age from the empirical relation and the one from the SED fitting for G4, giving a (mass-weighted) stellar age of $1-2$ Gyr. 
A region corresponding to the stellar age between 300 and 800 Myr where most post-starburst galaxies would fall, is marked according to \citet{Belli2019}. 
Two other different definitions of the quiescent region in the $UVJ$ color space are plotted: the dashed-dotted line is the definition suggested by \citet{Williams2009}, and the red solid line represents the criteria by \citet{Leja2019b} for $\log(sSFR/{\rm yr}^{-1}) = -10.5$. 
The dust attenuation direction and the amount are indicated as the arrow for $A_{V} = 0.4$.
We show two simple toy models of $UVJ$ color trajectory assuming the $\tau$-model SFH, for different $e$-folding times, 100 Myr (dark blue solid, fast track) and 1 Gyr (dark red dotted, slow track) using \textsc{fsps-py} following \citet{Belli2019} (see Section~\ref{sec:Qpath} for details).
\label{fig:uvj}}
\end{figure}

We investigate whether G4 also satisfies the selection criteria of quiescent galaxies using rest-frame $U$, $V$, $J$ magnitudes, based on the best fit from \textsc{bagpipes} (non-parametric SFH).
The $U-V$ and $V-J$ colors (hereafter $UVJ$ colors) are widely used to distinguish quiescent galaxies from reddened dusty galaxies and star-forming galaxies as originally demonstrated by \citet{Labbe2005} and \citet{Wuyts2007}; star-forming galaxies have bluer $U-V$ colors than the other two classes, and dusty star-forming galaxies tend to have redder $V-J$ colors than their quiescent counterpart.
This idea has been further supported by simulations for $z\sim2$ galaxies (e.g., \citealt{Donnari2019}, but see \citealt{Akins2022}).
The $UVJ$ diagnostics become less reliable for $z>3$ galaxies with decreasing purity both indicated by observations (e.g., \citealt{Merlin2018, DEugenio2020, Forrest2020}) and simulations \citep{Lustig2021, Lustig2023}.
Nevertheless, the $UVJ$ color selection works fairly well to select quiescent galaxies at $z\sim2$ observationally.

We adopt the $UVJ$ color selection of \citet{Whitaker2011} for $2<z<3.5$ galaxies which are shown as a black diagonal line and a dashed line in Figure~\ref{fig:uvj}. 
The diagonal black line separates the plane into two regions of quiescent and star-forming galaxies from their definition.
Additional criteria of $U-V >1.2$ and $V-J<1.4$ (dashed line) distinguish unobscured and dusty star-forming galaxies, respectively.
G4 is classified as a quiescent galaxy from these criteria.
The $UVJ$ colors based on \textsc{eazy-py} (filled square in Figure~\ref{fig:uvj}) also show a good agreement. 
In Appendix~\ref{app:sed}, we also list the best-fit $UVJ$ colors obtained from the best-fit SED without fixing redshift. 
All but one SED best-fit results give the rest-frame $UVJ$ colors of the quiescent population. 
One exceptional case is in the \textsc{bagpipes} run with a flat redshift prior using double-power law SFH, where the convergence was bad to reliably constrain the UVJ colors with large error bars (Table~\ref{tab:sedzphot}).

The $UVJ$ colors of G4 also satisfy two other different quiescent galaxy criteria proposed by \citet{Williams2009} and \citet{Leja2019b}.
\citet{Williams2009} proposed a slightly different definition as a quiescence ridge line, $U-V >1.3$, $V-J<1.6$ and $(U-V) > 0.88 \times (V-J) + 0.49$ for $z\sim2$ galaxies (a dashed-dotted line in Figure~\ref{fig:uvj}).
The solid red line in Figure~\ref{fig:uvj} is from \citet{Leja2019b} for $\log(sSFR/{\rm yr^{-1}}) = -10.5$, in which the sample purity of quiescent galaxies is high enough ($\gtrsim92\%$) in the $UVJ$ diagram using the 3D-HST galaxies.
There is a tendency for galaxies with lower sSFR to be placed on the top left corners, more distantly perpendicular to the QG-SFG dividing line (thick solid black line in Figure~\ref{fig:uvj}).
However, the sSFR distribution saturates beyond this sSFR limit such that $UVJ$ colors no longer give any further constraints.

Given that G4's $UVJ$ colors satisfy the quiescent criteria, we use this to infer the stellar age of the galaxy and compare it with the SED best fit.
Calibrated with their deep spectroscopic observations, \citet{Belli2019} found a good correlation with the (median) stellar age (for $>300$ Myr) and the $UVJ$ colors for $1.5<z<2.5$ quiescent galaxies and the age distribution is along the diagonal quiescent ridge line.
\citet{Belli2019} introduced a rotated coordinate system to calculate the age, 
\begin{eqnarray}
S_Q =0.75(V -J)+0.66(U -V) \nonumber \\
C_Q =-0.66(V -J)+0.75(U-V),
\end{eqnarray}
and the following median stellar age ($t_{50}$) can be estimated by $\log(t_{50}/\,{\rm yr})=7.03+1.12\cdot S_Q$.
This age-color trend was also corroborated by \citep{Carnall2020} based on the broad-band photometric data points for $2<z<5$ quiescent galaxies.

The mass-weighted age from the SED matches well with the one from the empirical relation.
The color of G4's symbol in Figure~\ref{fig:uvj} represents the mass-weighted age ($\sim1.9$ Gyr) based on the SED fitting using \textsc{bagpipes} (non-parametric SFH), as listed in Table~\ref{tab:sedbestfit}.
The small grey dots in the background show $\log(M_{\star}/M_{\odot})>10.5$ galaxies at the redshift range of $2.0<z<2.5$ from the COSMOS2020 catalog \citep{Weaver2020} by taking the rest-frame $UVJ$ colors from their $\textsc{eazy-py}$ fit.
The number density of them is also plotted as contours.
For visual comparison, we color-code the empirical age-color relation proposed in \citet{Belli2019} using the COSMOS2020 galaxies within the quiescent region.
The region marked in a blue box indicates the age range between $300<{\rm age/Myr}<800$, where post-starburst galaxies are expected to place according to \citet{Belli2019}.
Although we need spectroscopic confirmation to better constrain the stellar age, the SED-based mass-weighted ages (Table~\ref{tab:sedbestfit}, $\sim1-2$ Gyr) and empirically derived age ($\approx1.2$ Gyr) are broadly consistent within the errorbars. 

Typical ages of quiescent galaxies with spectroscopic observations at $z\sim2$ are observed to be $1-2$ Gyr, with a relatively large age spread \citep{Kriek2006, Kriek2009, Whitaker2013, vandeSande2013, Mendel2015, Onodera2015, Fumagalli2016, Belli2019, Stockmann2020,  Estrada-Carpenter2020}. 
This is also noticeable, although perhaps being much crude, by the inferred age and density distribution of COSMOS2020 galaxies in Figure~\ref{fig:uvj}; the density peak of the ``red cloud" is at $\sim1$ Gyr but there are many galaxies younger and older than this.
Considering the age distribution in the $UVJ$ diagram, the estimated stellar age of the G4 ($=1-2$ Gyr) appears to be reasonable and common at $z\sim2$.

\subsection{Dust-to-stellar mass ratio}\label{sec:dts}
\begin{figure}
\centering
\includegraphics[width=0.48\textwidth, bb=0 0 760 700]{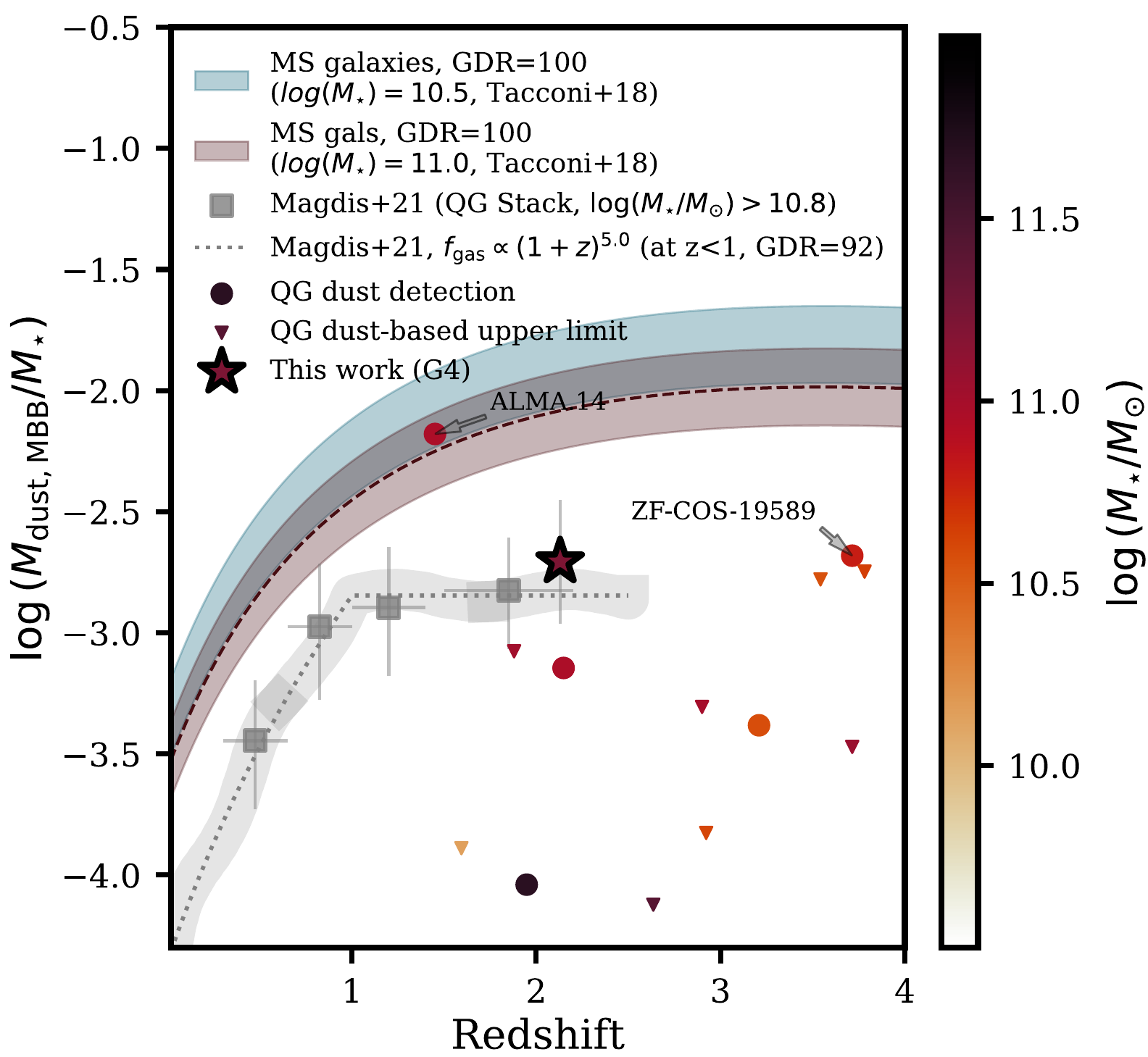}

\caption{Constraints on dust-to-stellar mass ratios as a function of redshift for $z>1$ quiescent galaxies. G4 is indicated by a star symbol. Sources from the literature are indicated as circles if they are detected in dust emission; otherwise, inverted triangles indicate upper limits (3$\sigma$). ALMA.14 and ZF-COS-19589, two galaxies with high dust-to-stellar mass ratios comparable to G4, are also labeled. The color of each source indicates the stellar mass of the galaxy. The stacking results by \citet{Magdis2021} for massive ($\log(M_{\star}/M_{\odot})>10.8$) quiescent galaxies at $z\lesssim2$ are shown as grey squares. Grey shaded region with a dotted curve is based on the functional form for the best-fit at $z<1$, $\mu_{\rm gas} (=M_{\rm gas}/M_{\star})\propto (1+z)^5$,  assuming a fixed gas-to-dust mass ratio of 92 and normalized at $z=1$ with $\mu_{\rm gas}$ = 8\%. For comparison to star-forming galaxies of similar mass, we calculated dust-to-stellar mass ratios using the scaling relation suggested by \citet{Tacconi2018},   assuming a gas-to-dust mass ratio (GDR) of 100 for $\log{(M_{\star}/M_{\odot})} = 11.0$ and $10.5$ (red and blue shaded bands, respectively; more massive galaxies have smaller gas (and dust) at fixed GDR.\label{fig:fdust_z}}
\end{figure}

We estimate the dust mass of G4 based on a single measurement at 2 mm, assuming a modified black body with
\begin{equation}
 T_{\rm d}=23.5\,{\rm K},\; \beta =2.08,\; \kappa_{\rm abs} (250\mu {\rm m}) = 4.0\,{\rm cm^2\, g^{-1}}
\end{equation}
where $\beta$ is the dust emissivity index, and $\kappa_{\rm abs}$ is the grain absorption cross section per unit mass, following the discussion in \citet{Bianchi2013} and \citet{Berta2016}.
The lower dust temperature of a quiescent galaxy is indicated by \citet{Magdis2021}, which performed the stacking analysis\footnote{We note that they excluded the 24$\mu$m detection sources above $\sim$45 $\mu$Jy (3$\sigma$) and G4 is below the limit, satisfying their selection criteria.} of the SED and obtained the best fit using \citet{Draine2007} templates.
The adopted value of 23.5 K is the value for $z\sim2$ quiescent galaxies\footnote{\citet{Magdis2021} noted a potential evolution in the dust temperature giving higher dust temperature with increasing redshift. 
If we take the mean value of 21 K estimated for $z=0-2$, the dust mass of G4 becomes slightly higher by 0.08 dex, giving $\log(M_{\rm d}/M_{\odot}) = 8.59 \pm 0.24$.}. 
The estimated dust mass of G4 is $\log(M_{\rm d}/M_{\odot}) = 8.51 \pm 0.24$ where the error takes into account the photometric error only. 

We note two caveats in our dust mass estimate: first, the inferred dust mass is a function of the assumed dust temperature, with higher $T_{\rm d}$ corresponding to lower dust masses given an observed flux density.  
The dust-to-stellar mass ratio for G4, assuming $T_{\rm d} = 23.5$ K, is 
${\rm log}(M_{\rm d}/M_{\ast}) = -2.71\pm 0.26$; if one instead assumes  
$T_{\rm d} = 25$ K, often adopted for normal star-forming galaxies \citep{Scoville2016}, the dust-to-stellar mass ratio reduces to ${\rm log}(M_{\rm d}/M_{\ast}) = -2.75$  --  still at the high end of the range measured for quiescent galaxies ($<-3$); see also Figure~\ref{fig:fdust_z}.
Second, the modified black body assumption we have used is a conservative estimate of dust mass. It is known that it gives systematically lower values compared to estimates based on a fully-sampled SED fitted with \citet{Draine2007} models. 
The current assumption of the $\beta$ and the corresponding $\kappa$ values mitigates this discrepancy \citep{Bianchi2013, Berta2016}.
The dust mass of G4 could become even higher if we adopt \citet{Draine2007} and once the SED is well sampled; if this is the case, it will strengthen the dust-rich nature of G4.
As further support of our estimated dust mass, we note that the dust mass derived from \textsc{cigale}, where we assumed the \citet{Draine2014} model (an updated model from \citealt{Draine2007}),  is $\log(M_{\rm d}/M_{\odot}) = 8.62\pm0.40$, consistent with our measurement assuming the modified black-body.

In Figure~\ref{fig:fdust_z}, we show the dust-to-stellar mass ratio of G4 ($\log({M_{\rm d}/M_{\star}}) = -2.71 \pm 0.26$) based on the above calculation and stellar mass from the SED fitting using \textsc{bagpipes} (non-parametric SFH).
For comparison, we plot the stacking results of quiescent galaxies with $\langle \log(M_{\star}/M_{\odot})\rangle \approx 11.0$ and $0 < z < 2$ from \citet{Magdis2021} (grey shaded curve and square data points after factoring in $\approx0.21$ dex for converting Salpeter (\citealt{Salpeter1955}) IMF to Chabrier IMF in stellar mass).

The $M_{\rm d}/M_{\star}$ value for G4 is comparable with those based on stacking analysis (\citealt{Man2016, Hayashi2018, Gobat2018, Magdis2021, Blanquez-Sese2023}) and higher than most of the individual gravitationally-lensed QGs \citep{Whitaker2021, Caliendo2021} (Figure~\ref{fig:fdust_z}).
\citet{Whitaker2021} put an order of magnitude lower limit of dust-to-stellar mass ratio ($<-4$ dex) for non-detected sources, although one must be cautious interpreting non-detections for more spatially-extended lensed sources \citep{Gobat2022}.  
Individual dust measurements and upper limits of non-lensed sources from \citet{Hayashi2018, Suzuki2022, Kalita2021} are also shown using the same modified blackbody model assumptions for $\beta$, $\kappa_{\rm abs}$, and $T_{\rm d}$ in Eq.(2). 
Most individual data points have dust-to-stellar mass ratios at least a factor of 3 lower than G4; the exceptions 
 are ALMA.14 ($z=1.4$; \citealt{Hayashi2018}) and ZF-COS-19589 ($z=3.7$; \citealt{Suzuki2022}), which have inferred dust-to-stellar mass ratios comparable to or even higher than G4 (labeled in Figure~\ref{fig:fdust_z}.)

ALMA.14 is located at the outer edge of the core region of a galaxy cluster at $z=1.4$\footnote{Note that we checked whether two gas(dust)-rich `quiescent' galaxies in \citet{Hayashi2018} and \citet{Rudnick2017} in clusters at $z=1.46$ and $z=1.62$ satisfy the quiescent criteria described in Section~\ref{sec:uvj}, and they do not satisfy the \citet{Leja2019b} criteria. The  \citet{Rudnick2017} galaxy lies outside of the \citet{Williams2009} definition of quiescence, and hence we excluded it from our analysis and from Table~\ref{tab:otherQG}.}. 
It is detected in both \cotwo\ and \oii\ (narrow-band); while it has a low ${\rm SFR} \sim 3$ M$_{\odot}$ yr$^{-1}$ based on an SED fit to the optical/near-IR photometry (see Table~\ref{tab:otherQG}),  this estimate may significantly under-estimate the actual SFR, as we have discussed (Section~\ref{sec:sedsfr}). 
We recall that G4's best-fit SFR using the same star formation history parametrization ($\tau$ model) and SED fitting code (\textsc{fast++}) suggested SFR$\sim 0.01$ M$_{\odot}$ yr$^{-1}$. Given the CO and \oii\ detections of ALMA.14, its status as a QG may be questionable. 

ZF-COS-19589 has an inferred SFR comparable to our best estimate SFR of G4 (Section~\ref{sec:sedsfr}) based on a dust-corrected H$\beta$ luminosity: $10^{+28.5}_{-12.9}$ $M_{\odot}$ yr$^{-1}$; \citet{Schreiber2018b}. 
Its SFR
inferred from an SED fit is $0.00^{+4.41}_{-0.00}\; M_{\odot}\, {\rm yr^{-1}}$.
\citet{Suzuki2021} estimated the $SFR_{\rm IR}$, extrapolated from a single continuum detection at an observed wavelength of 870$\mu$m giving a range of $6-33$ $M_{\odot}$ yr$^{-1}$ depending on the assumed dust temperatures (between 20 K and 40 K). These inferences suggest that
ZF-COS-19589 may have properties similar to G4, with M$_{\ast}$ smaller by
$\sim0.3-0.4$ dex but with a comparable deviation ($-1.2$ dex) from the star-forming main-sequence for its redshift.

In Figure~\ref{fig:fdust_z}, we also plot the dust-to-stellar mass constraints of the star-forming main-sequence galaxies for different stellar mass bins of $\log{(M_{\star}/M_{\odot})}=11.0$ and $10.5$.
We used \citet{Tacconi2018} relation to get gas fraction and converted it to dust-to-stellar mass ratio assuming a fixed gas-to-dust mass ratio (GDR) of 100. 
The shaded region corresponds to $\pm0.3$ dex scatter of the main sequence.
The adopted GDR is reasonable for these two massive populations with high metallicity e.g., \citet{Remy-Ruyer2014}. 
For a higher GDR, the shaded bands would go lower and vice versa.

If the Kennicutt-Schmidt relation is also applicable to this galaxy, we expect the gas-to-dust mass ratio of G4 should be higher than 100 to explain the observed dust-to-stellar mass ratio. 
A high gas-to-dust mass ratio for lensed (less massive) quiescent galaxies at $z\sim2-3$ was also advocated in \citet{Whitaker2021} based on \textsc{simba} simulation. 
To put it in a different way, the galaxy's depletion time scale is long ($\gtrsim$2.5 Gyr) at $z\sim2$ with the estimated SFR and observed dust at fixed gas-to-dust mass ratio.
We will explore the dust-rich nature of the galaxy in the next section (Section~\ref{sec:dustrichness}).

Taking these all into account, G4 is highly dust-rich with a dust-to-stellar mass ratio of $\log(M_{\rm d}/M_{\star})=-2.71\pm0.26$, compared to other high redshift quiescent galaxies reported in the literature, except for two and the stacked results.
Considering its current star-formation rate at a low level ($\log(sSFR/{\rm yr})\lesssim-10.2$), unless there is a process igniting star formation at later times, it is likely G4 will remain gas(dust)-rich while being ``quenched" for a long period of time ($\gtrsim$2.5 Gyr).

\subsection{Morphology}\label{sec:morphology}

The HST composite (Figure~\ref{fig:hst}) image shows that G4 has a central core-like component and an extended disk. 
Here, we quantify G4's morphology by fitting the 2D light distribution with the S\'{e}rsic profile based on the F140W map for later discussion.

We use \texttt{statmorph} \citep{Rodriguez-Gomez2019}\footnote{\url{https://statmorph.readthedocs.io/en/latest/overview.html}} and \texttt{galfit} (ver3.0; \citealt{galfit_orig,galfit_v3}).
We fit the magnitude ($m$), half-light radius ($r_e$), S\'{e}rsic index ($n$), axis ratio ($q$), position angle (PA), and the central position. The initial guess parameters in the \texttt{galfit} are taken from the \texttt{statmorph} result.
The fit gives $n=1.66\pm 0.03$ (\texttt{statmorph}), where the error is estimated based on the bootstrapping after 1000 realization, and $n=1.66\pm0.11$ (\texttt{galfit}), respectively. The visual inspection of the residual images for both fits suggests the fit is reasonably good.

We also obtained Gini \citep{Abraham2003} and $M_{20}$ \citep{Lotz2004} non-parametric values from \texttt{statmorph}. They are $0.52\pm0.01$ and $-1.85\pm0.03$, respectively.
These values locate the galaxy close to the borderline between Sb/Sbc and E/S0/Sa classification based on \citet{Lotz2008}.

\section{A dust-rich quiescent galaxy in distant universe}\label{sec:discussion}
The available opt/NIR/MIR photometry and best-fit $UVJ$ colors (although these two are not independent measurements) suggest that G4 is a quiescent galaxy with low sSFR and mass-weighted age of $\approx1-2$ Gyr at $z\approx2$.
Meanwhile, ALMA observations reveal that the galaxy has a high dust-to-stellar mass ratio, comparable to the stacked results which are (at least threefold) higher than most of the quiescent galaxies individually studied.
In this section, we take advantage of these findings and compile available measurements in the literature to address the implication of G4's properties in the context of quenching at high redshift.

\subsection{Age-$\Delta$MS relation: passive evolution at $z>1$?}\label{sec:age-dms}

\begin{figure}
\centering
\includegraphics[width=0.47\textwidth, bb=0 0 930 830]{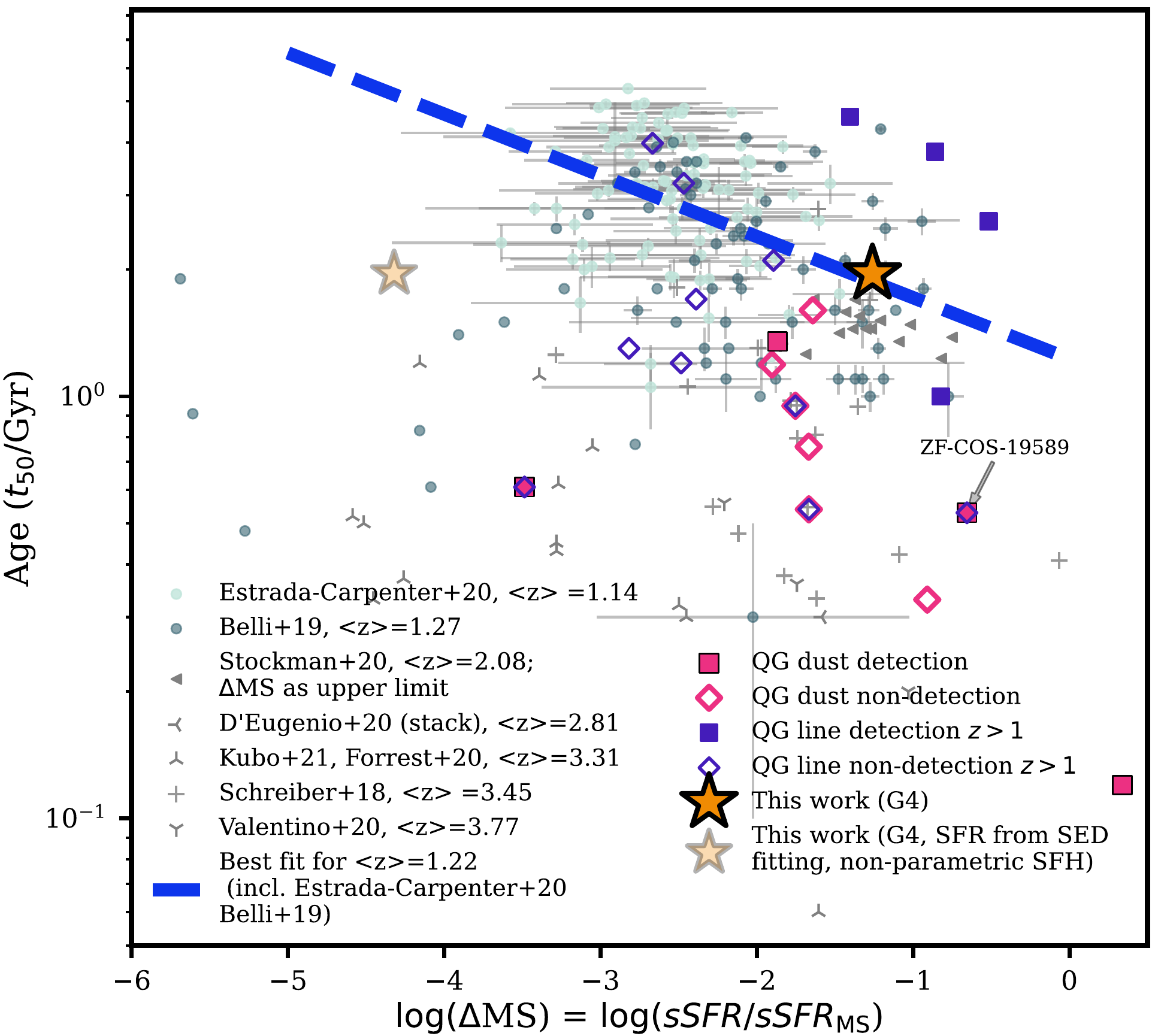}
\caption{Age (mass-weighted) distribution as a function of the offset from the main sequence. G4 is plotted as star symbols, using different SFR estimates (24$\mu$m flux scaling and SED); the fainter one is the best-fit SED using \textsc{bagpipes} (non-parametric SFH) and see Section~\ref{sec:sedsfr} for detailed discussions. 
We use the main-sequence definition from \citet{Speagle2014}. 
Literature values are from \citet{Schreiber2018b, Belli2019, Estrada-Carpenter2020, Stockmann2020, DEugenio2020, Valentino2020a, Kubo2021} for $z\sim1-4$ galaxies with constraints of the mass-weighted age, stellar mass, and star-formation rate. 
Here, we use SFR values from the SED fitting, except for two cases (ZF-COS-19589, \citealt{DEugenio2020}); for ZF-COS-19589, the 870$\mu$m flux is used to calculate SFR(IR) assuming $T_{\rm d} = 40$ K taken from \citealt{Suzuki2021};
\citet{DEugenio2020} performed stacking analysis and we took the dust-corrected SFR from the (stacked) [OII] line flux.
The thick blue dashed line shows our best fit of age-$\Delta$MS relation for $z\sim1$ quiescent galaxies using \citet{Belli2019} and \citet{Estrada-Carpenter2020} good quality samples (see Section~\ref{sec:age-dms}).
High-z ($z>1$) quiescent galaxies with dust/CO/[C I] observations are also plotted, taken from the various literature (see Section~\ref{sec:dustrichness} and Table~\ref{tab:otherQG}). 
\label{fig:age_dms}}
\end{figure}

If galaxies evolve passively after the quenching, a negative correlation between the age and the distance from the main sequence is expected.
With this idea, we first search for a signature of passive evolution at $z\gtrsim1$ in the age-$\Delta$MS relation.
We gather the measurements of $z\gtrsim1$ quiescent galaxies in the literature from \citet{Estrada-Carpenter2020, Belli2019} for $z\sim1$, \citet{Stockmann2020} for $z\sim2$, and \citet{Schreiber2018b, Valentino2020a, Forrest2020, DEugenio2020, Kubo2021} for $z>3$ sources, all of which are plotted in Figure~\ref{fig:age_dms}.

First, we investigate $z\sim1$ quiescent galaxies based on \citet{Estrada-Carpenter2020} and \citet{Belli2019} which provide 177 quiescent galaxies in total.
The age ($z_{50}$), stellar mass, and SFR measurements in these studies are estimated in a similar way, alleviating the systematic errors between the two.
We choose a subset of good-quality data sets with the following criteria: galaxies having $M_{\star}$ errors within 0.2 dex with the stellar mass of $\log(M_{\star}/M_{\odot})>10.5$ and SFR errors within 0.6 dex.
The adopted errors would sufficiently incorporate the potential systematic uncertainties in the different SED codes, as demonstrated by \citet{Pacifici2023}. 
The final number used in the fitting is 147 in total.
Galaxies with error bars in Figure~\ref{fig:age_dms} show the subset and are used in the following fitting procedure.

With the following functional form, we perform the Bayesian Markov chain Monte Carlo fitting using \texttt{PyMC} \citep{pymc}, taking into account the errors :
\begin{equation}
\log({\rm Age/Gyr}) =  A * \log(\Delta {\rm MS}) + B
\end{equation}
We set the uniform prior around the best fit from the orthogonal distance regression ($A=-0.26, B=-0.14$) with $dA = [1.5A,0.5A], dB = [2B, -2B]$.
The best-fit is A = -0.14 (confidence interval (CI) [3\%, 97\%] = [-0.19,-0.10]), B = 0.09 (CI = [-0.02,0.20]).
The Spearman coefficient (with the \textit{good quality} data) is $-0.43$ (thus negative) with $p$-value of 7$e$-8.
Therefore, we confirm that the anticorrelation of mass-weighted stellar age and the deviation from the main sequence exists at $z\sim1$, indicating passive evolution.
The best fit is shown as a dashed blue line in Figure~\ref{fig:age_dms}.


Compared to the situation at $z\sim1$, confirming the existence of age-$\Delta$MS relation at $z\gtrsim2$ is largely limited by the data. 
At $z\sim2$, \citet{Stockmann2020} provided mostly the upper limit constraints of SFR from the SED fitting.
At least for those detected in 24$\mu$m (table 2 in their paper), they are located close to G4 (i.e., $\log({sSFR})\simeq-10.2$ and $\log({\rm age})$ of $\simeq$9.2). 
Higher redshift quiescent galaxies at $z\gtrsim3$ (\citealt{Schreiber2018b, Valentino2020a, Forrest2020, DEugenio2020, Kubo2021}) are typically younger than G4 because the age of the universe gets younger.
While there is a hint of negative correlation with lower normalization in $z\sim3$ quiescent galaxies from the visual inspection, substantiating the existence of the relation $z\gtrsim2$ (as evidence of passive evolution) requires more observational data.

For completeness, in Figure~\ref{fig:age_dms}, we mark high-$z$ quiescent galaxies with dust continuum and/or CO/[CI] line measurements by compiling all available information ($M_{\star}$, age, SFR, dust/CO/\cone constraints) from the following papers: \citet{Onodera2012, Bezanson2013, Bezanson2019, Newman2018a, Hayashi2017, Hayashi2018, Suess2017, Rudnick2017, Glazebrook2017, Belli2015, Belli2019, Belli2021, Caliendo2021, Gobat2022, Whitaker2021, Williams2021, Morishita2022, Suzuki2022, Kalita2021, Akhshik2023}. 
We provide a summary in Table~\ref{tab:otherQG}.
Although it is tentative, it seems that many galaxies detected in dust and/or CO/\cone\, are more scattered with respect to age-$\Delta$MS relation. 
We investigate this more by connecting the dust-to-stellar mass ratio together with age and sSFR in Section~\ref{sec:dustrichness}.

\begin{figure}
\centering
\includegraphics[width=0.47\textwidth, bb=0 0 1000 700]{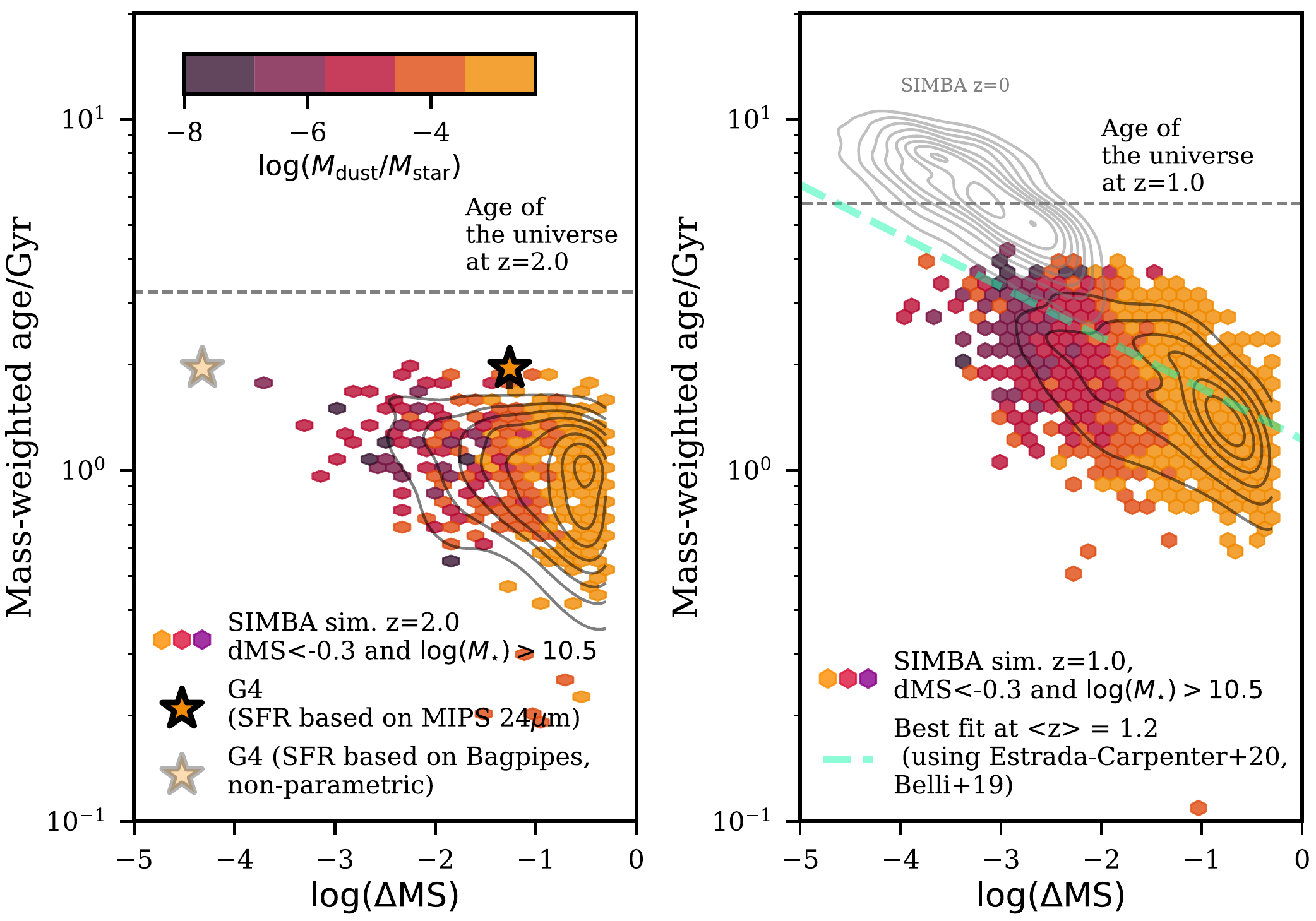}
\caption{Age-$\Delta$MS relation in \textsc{simba} simulations at $z=2.0$ (left) and $z=1.0$ (right). Data points are for massive ($\log{M_{\star}>10.5}$) quiescent galaxies (defined as $\log(\Delta{\rm MS}) <-0.3$) color-coded by the dust-to-stellar mass ratio. 
The distribution of $z=0$ quiescent galaxies are also shown in grey contours on the right panel.
We overlay G4's position in star symbols for the $z=2.0$ snapshot. 
The dashed line on the right panel is the best fit obtained from the observations of $z\sim1$ quiescent galaxies shown in Figure~\ref{fig:age_dms}. \label{fig:simba}}
\end{figure}

The anti-correlation at $z\sim1$ is also discernible in the cosmological simulation in \textsc{simba} simulations (\citealt{Dave2019}\footnote{\url{http://simba.roe.ac.uk/}}; Figure~\ref{fig:simba}), supporting our discovery based on the observational results.
\textsc{simba} includes an on-the-fly dust production and destruction model \citep{Li2019} that no other large simulation boxes ($\gtrsim (100\,h^{-1}$ Mpc)$^3$) currently allow. 
The results are taken from the flagship run snapshots at three redshift bins ($z=2.0, 1.0, 0.0$) with full \textsc{simba} physics in (100 $h^{-1}$ Mpc)$^3$ box using \texttt{CAESAR}\footnote{\url{https://github.com/dnarayanan/caesar}}.
Figure~\ref{fig:simba} shows the distribution of massive galaxies ($\log(M_{\star}/M_{\odot})>10.5$) below the main-sequence ($\log(\Delta{\rm MS}) <-0.3$) on the age - $\log(\Delta$MS) plane.

Our best-fit result of $z\sim1$ quiescent galaxies (dashed line) is shown in the $z\sim1$ snapshot (Figure~\ref{fig:simba} right), which is comparable to the simulation result.
The consistency in the observation and simulation at $z=1.0$ and the evolution down to $z=0$ (shown in grey contours) suggest that $\Delta$MS-age relation is likely well established up to $z\sim1$ for quiescent galaxies.

The correlation at $z=2.0$ is not clearly recognizable in the simulation (but a tail exists).
\citet{Mendel2015}, with their spectroscopic campaigns (VIRIAL survey), demonstrated that stellar ages of $z\lesssim1$ quiescent galaxies are reasonably explained by passive evolution while at $z\gtrsim2$, galaxies' pathways after the quenching (or while being quenched) and the evolution may not be simply a ``passive" evolution.
The lack of age-$\Delta$MS correlation at $z\sim2$ in the simulation perhaps corroborates the finding from the VIRIAL survey.

Meanwhile, \textsc{simba} simulations hardly find galaxies displaying similar properties of G4 for its given $\Delta$MS and age and considering the associated uncertainties (see black contours in the left panel of Figure~\ref{fig:simba}).
The distribution of G4 and available $z\sim2$ quiescent galaxies highly encourages more observations to check if there is any difference between the observation and simulation at $z\gtrsim 2$ and whether G4 is instead a rare population in the $\Delta$MS-age relation.
Finally, the SFR from 24$\mu$m shows a \textit{better} agreement with the distribution of simulated galaxies, providing indirect support for our choice of the SFR.

\subsection{Long depletion time scale}\label{sec:dustrichness}
If galaxy quenching requires the removal/consumption of gas and dust, we can conjecture an additional correlation (that may not necessarily be linear) with the gas and dust content with the deviation from the main sequence (or sSFR) and age. 
Initial reduction/consumption of gas and dust (just after quenching) would have an imprint on the strength of the initial quenching mechanism, while the following evolution would give us a hint on additional mechanisms to remove/consume/supply gas and dust out of/to the galaxy.

Such a possibility is hinted at in the \textsc{simba} galaxies as shown in Figure~\ref{fig:simba} where the colors represent the dust-to-stellar mass ratio at given $\Delta$MS and age; we find a decrease of $M_{\rm d}/M_{\star}$ along the ``passive evolutionary" sequence identified in Section~\ref{sec:age-dms}.
Observations also found such a trend between $M_{\rm d}/M_{\star}$ and sSFR in the local galaxies (e.g., \citealt{Cortese2012, DeVis2017a}).
All these above lead us to connect the observed sSFR and age with the observed dust.

\subsubsection{Dust depletion time scale}\label{sec:longdepl}
\begin{figure}
\centering
\includegraphics[width=0.48\textwidth, bb=0 0 1000 700]{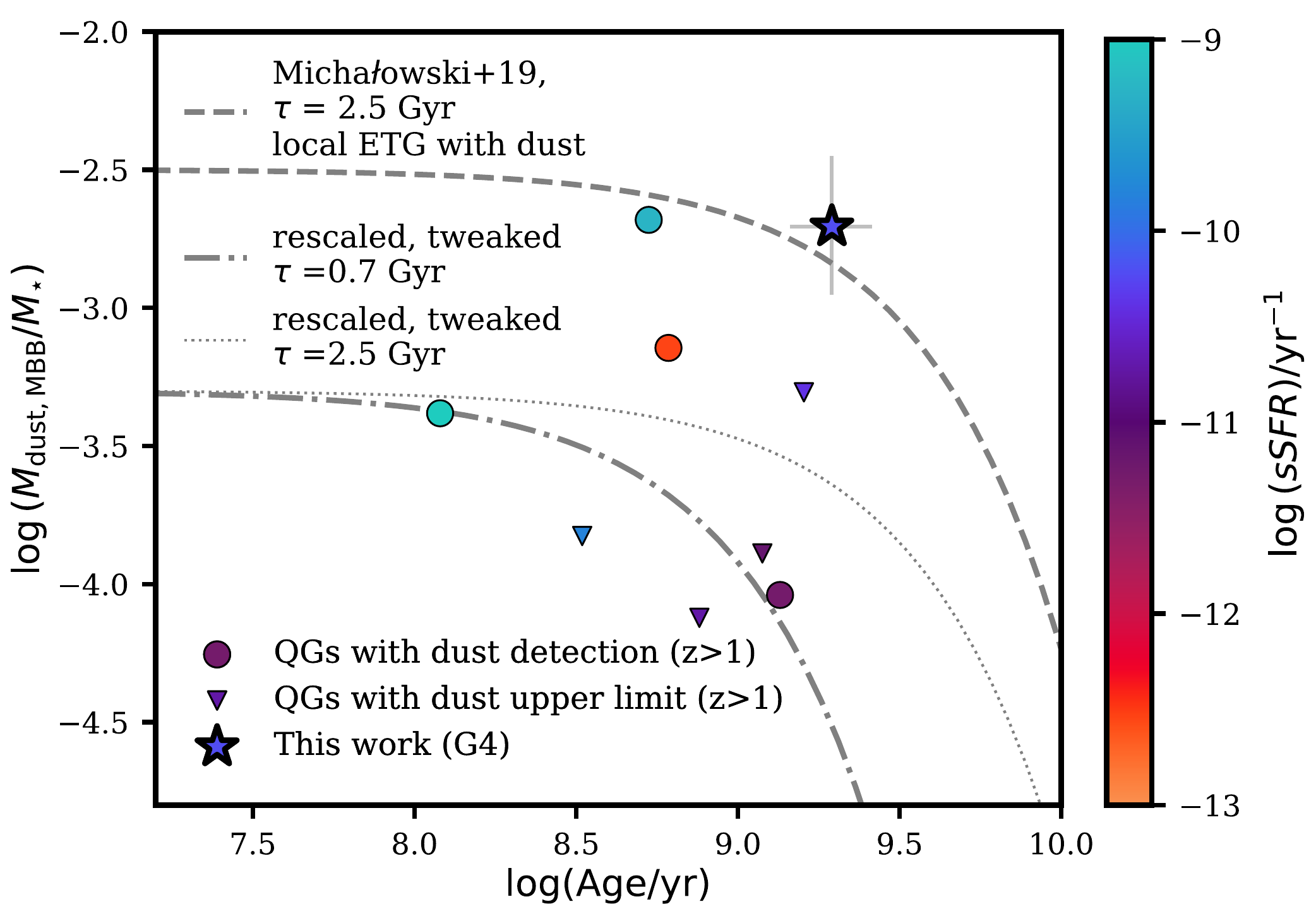}
\caption{Dust-to-stellar mass ratio as a function of mass-weighted age. G4 is plotted as a star color-coded with its sSFR. We also plot other $z>1$ quiescent galaxies with dust constraints from the literature (detection for circles, 3$\sigma$ upper limit with upside-down triangles for non-detection; \citealt{Gobat2022, Whitaker2021, Suzuki2022, Kalita2021}). The three lines are inspired by \citet{Michalowski2019}: dashed line is the fit based on the local Herschel selected early-type galaxies, while the other two lines are tweaked and rescaled assuming $\tau = 0.7$ Gyr (dashed-dotted) and $2.5$ Gyr (dotted) to guide our eye for dust-poor high-redshift quiescent galaxies. The colors represent the specific star-formation rate of the galaxies.\label{fig:dts_age}}
\end{figure}

With the data set compiled in Table~\ref{tab:otherQG}, we show the $\log(M_{\rm d}/M_{\star})$ as a function of $\log({\rm Age})$ in Figure~\ref{fig:dts_age}.
Individual data points are from this work (G4), \citet{Gobat2022}\footnote{We take the point source assumption, which is comparable to the original report in \citet{Whitaker2021}. According to \citet{Gobat2022}, the dust mass (as upper limit) could be $\times$6 higher for extended source assumption. Nonetheless, G4's dust-to-stellar mass ratio would still be higher than any of those sources by $\gtrsim0.3$ dex.}, \citet{Suzuki2022}, and \citet{Kalita2021}, giving nine sources in total.
From this figure, the dust-richness of G4 at a given stellar age is a noticeable feature. 
It is also conceivable that there are two different subgroups in this plane: five out of nine galaxies display coherent decreasing sSFR with their age, while the remaining four do not with higher dust-to-stellar mass ratios including G4. 
In the following discussion, we call the former dust-poor quiescent galaxy and the latter dust-rich quiescent galaxy.

We compare G4 with these high redshift quiescent galaxies by taking the functional form ($\frac{M_{\rm d}}{M_{\star}} = A \exp({\rm -age/\tau})$) described in \citet{Michalowski2019}.
\citet{Michalowski2019} used the Herschel-detected early-type galaxies (ETG) in the local universe ($z<0.1$), and measured the $e$-folding time of $\tau = (2.5\pm0.4)$ Gyr.
This is shown as a dashed line in Figure~\ref{fig:dts_age}.

G4 is well aligned with the measurement of the local dust-rich ETG.
The other (three) dust-rich quiescent galaxies are also close to this relation. They are MRG-2129 ($z=2.1$; \citealt{Whitaker2021, Man2021, Gobat2022, Morishita2022}) --- known to be rotation dominated (\citealt{Toft2017, Newman2018b}) and host AGN \citep{Man2021} ---, 
Galaxy-D (z=2.9; \citealt{Kalita2021}) --- located in a group environment --- and ZF-COS-19589 ($z=3.7$ but uncertain redshift; \citealt{Schreiber2018b, Suzuki2022}).
The SFR levels are different by more than three orders of magnitude in these galaxies.
We also note that except for G4, dust emissions are captured at shorter wavelengths ($<400\,\mu$m) in the rest frame which is a subject of a larger uncertainty to get the true dust mass.
Having this cautionary note in mind, the two most dust-rich quiescent galaxies (G4 and ZF-COS-19589) are aligned with the local relation measured in \citet{Michalowski2019}.
Intriguingly, the stacked data from \citet{Blanquez-Sese2023} at $\langle z\rangle\sim1.5$ is located closer to G4 if we use the inferred stellar age using \citet{Belli2019} from their rest-frame $UVJ$ colors ($U-V$ = 1.73, $V-J$ = 1.05, weighted mean, giving log(age)=9.2), also being consistent with the trend of \citet{Michalowski2019} with longer depletion time.

\citet{Donevski2023a} explored 548 dusty quiescent galaxies in the hCOSMOS survey at $0.01<z<0.7$, finding a similar evolutionary trend in the spiral quiescent galaxies as observed in \citet{Michalowski2019}; elliptical quiescent galaxies show an almost flat trend with age, demonstrating that morphological type is an important factor of the scatter in $M_{\rm d}/M_{\star}$.
We recall the G4's morphology is a composite of disk and bulge structures (Section~\ref{sec:morphology}). 
Perhaps, G4 and other dust-rich quiescent galaxies could be the progenitors of dusty (spiral) quiescent galaxies which constitute $\sim10\%$ of the total quiescent population in \citet{Donevski2023a}, and of those in \citet{Michalowski2019}.
We may be witnessing the emergence of such (relatively rare) galaxy population observed at lower redshift.

If we rescale and match the functional form by hand for the remaining dust-poor quiescent galaxies, we estimate the dust-depletion time scale of $\tau\lesssim 0.7$ Gyr as shown as a dash-dotted line in Figure~\ref{fig:age_dms}.
\citet{Whitaker2021} explored the nature of high redshift quiescent galaxies without dust detection.
They showed a substantial drop (or a huge scatter) in the $M_{\rm d}/M_{\star}$ ratio at the (light-weighted)\footnote{In general, we would expect the mass-weighted age would be higher than this if there has been a recent star-formation activity.} age of $\sim0.3-0.5$ Gyr for $z\sim2$ galaxies (Figure~5 in their paper).
This corresponds to the (slightly longer) dust destruction time scale of theoretical expectations ($0.1-0.4$ Gyr, e.g., \citealt{Draine1979, Jones1994}).
They attributed the dust non-detection as an indication of extremely high gas-to-dust mass ratio ($\delta_{\rm GDR}$) and of the mechanisms requiring strong dust destruction and shutting down star formation -- and this may also hamper efficient grain growths.

Altogether, G4 may have experienced less efficient feedback (see also Section~\ref{sec:dustyield}) that would be in the form of dust destruction by the supernova (SN) reverse shock (e.g., \citealt{Nozawa2006}), thermal sputtering by hot electrons (e.g., \citealt{DeVis2017a, Galliano2021}) and astration (e.g., \citealt{Clark2015}) compared to dust-poor ones. 
Further, a relatively larger scattered distribution of dust-rich quiescent galaxies in Figure~\ref{fig:age_dms} could also mean some additional processes that play a role by postponing the full depletion of the dust in G4.

\subsection{Quenching path and mechanism to maintain quiescence}\label{sec:Qpath}
The dust-richness with only a moderate SFR of G4 is puzzling because of the large inferred amounts of gas (from dust) to potentially (re-)fuel star formation if it cools fast enough.
Unless there is a mechanism that hampers gas to form stars e.g., morphological quenching (\citealt{Martig2009}), it is difficult to explain simultaneously the quiescence and the dust-rich nature of G4.
We describe in the following that G4 cannot be simply explained with a fast-quenching mode unless there is an additional mechanism which \textit{aids} sustaining the dust-rich nature.

We revisit the toy model presented in \citet{Belli2019} using the python interface of \textsc{fsps} \citep{Conroy2009b}, \textsc{python-fsps} \citep{pyfsps}.
Synthetic galactic spectral energy distributions and the $UVJ$ trajectories are constructed using the exponentially declining SFH model for two different $e$-folding times of 100 Myr and 1 Gyr. 
The isochrones are based on \textsc{MIST} \citep{Choi2016, Dotter2016} and the spectral library is from \textsc{MILES} \citep{Falcon-Barroso2011}.
We used the dust attenuation law of \citet{Calzetti2000} and the old stellar population attenuation (\texttt{dust2}) being 0.4 (which corresponds to roughly $A_{\rm V}\approx0.4$), which would also incorporate the $A_{\rm V}$ values from the SED fitting (Table~\ref{tab:sedbestfit}). 
We fixed the metallicity as solar metallicity.
Two trajectories are shown in Figure~\ref{fig:uvj}.
We also show the expected offset with dust which we indicate as the arrow of $A_{\rm V}=0.4$.
The trajectories would depend on SFH and dust attenuation but show the overall trends of the different quenching time scales.
The information of G4's SFH before the quenching remains unconstrained (i.e., the best fit can be obtained in many parametric models as well as non-parametric SFH, see Section~\ref{sec:sed_model}).
Nevertheless, G4's colors and the expected age support that the galaxy has experienced rather a faster quenching episode based on the \textsc{fsps} model trajectories. 

The fraction of rapid quenching mode increases from $\sim4\%$ at $z=1.5$ to $\sim23\%$ at $z=2.2$ \citep{Park2022}. The number would go higher if one adds a less bursty post-starburst track.
\citet{Park2022} also found in the TNG100 simulation that rapid quenching in these galaxies is driven by AGN.
For half of the cases, gas-rich major mergers trigger the central starburst and make a compact remnant, although TNG100 underpredicts the fraction of fast-quenching phase galaxies than the observations.

In a classical picture where AGN plays a role, it accompanies the efficient removal/destruction of gas and dust. 
If the fast quenching mode was driven by AGN, G4 is not aligned with this picture because it still has a considerable amount of dust at a given stellar mass. 
It can be a moderate starburst or AGN that does not efficiently or entirely remove or destroy dust, and the remaining dust (and gas) is not cooled and collapsed fast enough to form stars, perhaps by the gravitational potential of a bulge component.

G4's dust-richness at a given sSFR and age and long dust-depletion times may indicate a morphological quenching \textit{together with} inefficient (or failed) feedback. 
G4's morphology (Section~\ref{sec:morphology}) is fitted with a moderately high S\'{e}rsic index, $n\approx1.6$.
\citet{Gobat2017} measured high dust content for $\langle z\rangle=1.7$ quiescent galaxies with high $n(=3.5\pm0.1)$ based on stacking analysis.
Although G4 has a somewhat lower S\'{e}rsic index, our results still hint at a compact bulge-like structure that could stabilize itself and hamper the fragmentation of cold gas to form stars (see also discussions in \citealt{Hjorth2014} which also proposed a need of morphological quenching for galaxies with excessive $M_{\rm d}$ at low SFR regime).

We infer the Toomre $Q$ parameter of G4 foreseeing future follow-up observations. 
For a stable disk, the ratio of rotation velocity and intrinsic gas dispersion is connected to a gas fraction ($f_{\rm gas}$) and Toomre Q parameter (\citealt{ForsterSchreiber2006, Genzel2011}):  
\begin{eqnarray}
\frac{v_{\rm rot}}{\sigma_0} = \frac{a}{f_{\rm gas} Q} 
\end{eqnarray}
MRG-2129 is one of the dust-rich quiescent galaxies that do not follow $M_{\rm d}/M_{\star}$ - age - sSFR relation (see Section~\ref{sec:longdepl}, Figure~\ref{fig:dts_age}).
This galaxy is known to be rotation dominated \citep{Toft2017, Newman2018b} with "high" $V_{\rm rot}/\sigma_0$ (stellar) value of $\sim2-3$.
If we assume that G4's disk is also rotationally supported ($V/\sigma = 2-3$ with $a=\sqrt{2}$, though the gas distribution can be different from the stellar component), its gas fraction ($f_{\rm gas}\approx 0.2 (20\%)$, converted from the observed $M_{\rm d}$ assuming GDR of 100) would suggest $Q \sim 2-3$.
Future stellar and gas kinematics studies will verify if G4 also has a stable disk further supporting the morphological quenching.

\subsection{High dust yield}\label{sec:dustyield}

In this last section, we briefly discuss the possible origins of the observed dust by quantifying dust yields. 
The discussion also aims to provide support for inefficient feedback acted on G4 and the need for additional mechanisms to explain the observed dust.

\begin{figure}
\centering
\includegraphics[width=0.48\textwidth, bb=0 0 860 600]{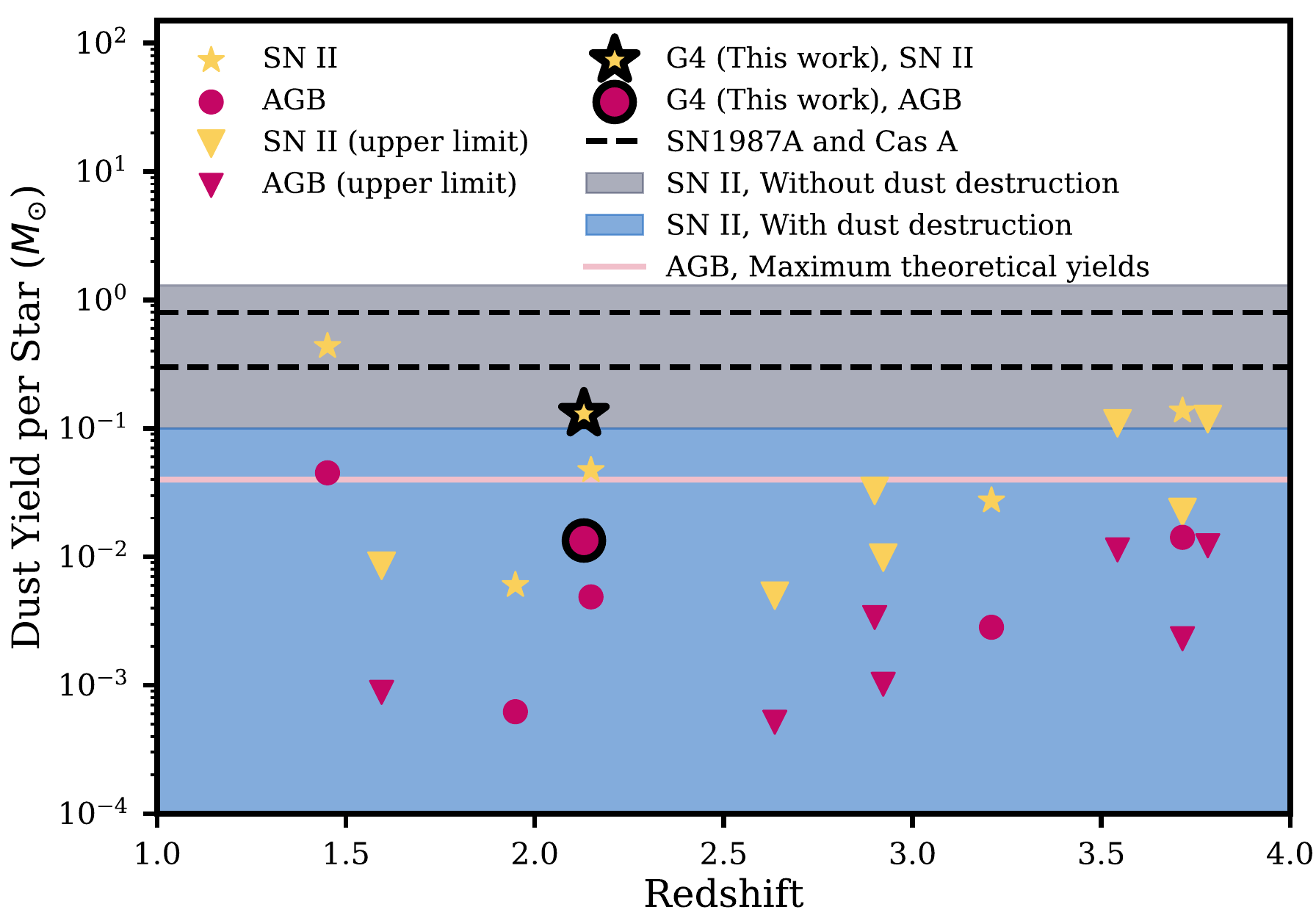}
\caption{Dust yield required for SNII and AGB stars to explain the observed dust in high redshift ($z>1$) quiescent galaxies including G4. The dust yield of SN II is in yellow star and for AGB in magenta circle for dust detected sources. Upper limits are shown as inverted triangles of each color.
Two dashed lines indicate the maximum dust yield range observed in SN 1987A and Cas A (0.3-0.8 $M_{\odot}$). The gray shaded region indicates the range of the dust yield expected from theory for SN II without dust destruction, while the blue area represents a more realistic yield range due to dust destruction ($\ll0.1\, M_{\odot}$). The pink line indicates the maximum theoretical yield possible for the AGB stars (0.04 $M_{\odot}$) but the observational results claim yields lower by more than an order of magnitude (see details in the main text). G4 requires almost the maximum dust yields from both SN II and AGB stars if grain growth is not taken into account. \label{fig:dust_yield}}
\end{figure}

We follow the calculation of \citet{Michalowski2010a} but with slightly different mass ranges when calculating the number of stars, $M_{\star} = 8-40\, M_{\odot}$ for Type II SN (hereafter SN II), and $M_{\star} = 1.5-8\, M_{\odot}$ for AGB stars.
With our estimated stellar mass from \textsc{bagpipes} (non-parametric SFH), one SN II should produce $\approx\,0.13\, M_{\odot}$ of dust, while $\approx0.013\, M_{\odot}$ for AGB star, to explain the observed dust mass ($3.2\times10^8\, M_{\odot}$).

Both values are close to the high end of the dust yield required for SN II and AGB stars.
For example, the maximum dust yields of the observed SN II are reported to be $0.45-0.8\, M_{\odot}$ for SN1987A \citep{Dwek2015, Matsuura2015}, $0.3-0.5\, M_{\odot}$ for Cas A \citep{DeLooze2017}, and $0.03-0.23\, M_{\odot}$ for Crab \citep{Gomez2012, Temim2013, DeLooze2019}.
However, these high values are for freshly formed dust.
Theoretical yield in AGB stars is 0.04 $M_{\odot}$ (e.g., \citealt{Morgan2003, Ferrarotti2006, Ventura2012, Nanni2013, Nanni2014, Schneider2014}), which is only valid for a very narrow range of the stellar mass.
The net yields of either origin are expected to be at least an order of magnitude lower if we take into account the destruction of the reverse shock for the SNe (e.g., \citealt{Todini2001, Nozawa2003, Nozawa2006, Bocchio2016}; blue shaded area in Figure~\ref{fig:dust_yield}) and the observational result of the AGB stars (e.g., \citealt{Rowlands2012}).

\citet{Donevski2023a} explored \textsc{simba} simulation and the chemical model of \citet{Nanni2020}, and offered the need for prolonged dust grain growth in dusty quiescent galaxies at intermediate redshifts.
The mechanism allowing efficient grain growth for high redshift sources is yet fully understood, with a limited amount of data. 
Exploring this is beyond the scope of this paper and the grain growth remains a possibility of the observed dust in quiescent galaxies.

On the other hand, \citet{Gobat2017} and \citet{Belli2017b} opened a possibility of the external origin of dust and gas for high redshift quiescent galaxies.
Considering the downsizing in the concordance model, the majority of massive quiescent galaxies in the early universe are expected to be located in dense environments.
At the time when the accretion rate from the large-scale structure is higher ($z=1-3$), there may be also a higher chance of a rejuvenation event for a (temporarily) ``quenched galaxy" no matter what quenching mechanism was playing a role initially.
In this regard, \citet{Tacchella2022} found an indication higher fraction of rejuvenated galaxies ($\approx30\%$) in a massive halo ($\log(M_{\rm h}/M_{\odot})>13.0$) compared to a lower halo ($\approx4\%$) for $z\sim0.8$ quiescent galaxies.

Based on the stellar-to-halo mass relation \citep{Moster2010}, the massive nature of G4 hints at the halo mass is $\log(M_h/M_{\odot})\sim 12.9$, which indicates the galaxy is located in a ``group-like" environment.
Confirming the overdensity and the association of G4 with star-forming galaxies (BX610 and G1) will further strengthen this idea\footnote{The stellar masses of BX610 and G1 are $\log{(M_{\star}/M_{\odot})} \sim 11.1$ and $10.4$ using the same method presented here. If G4 is at the same redshift and combining each inferred halo mass, the total halo mass would be $\log{(M_{\rm halo}/M_{\odot})}\sim 13.2$.}.
We estimate up to a third of the observed dust can be originated externally based on a crude calculation following the major merger rates for galaxies in the CANDELS field (\citet{Duncan2019}), which is 0.3-0.4 Gyr$^{-1}$ for $z\sim2$ $\log{(M_{\star}/M_{\odot})}>10.3$ galaxies.
Still, it would require a high dust survival rate with no destruction, likely assisted by efficient grain growth.
Follow-up observation of stellar and gas kinematics and their alignment could give us a hint of the external origin, as indicated in local fast rotators (\citealt{Davis2011, Davis2013}).

Taking these all together, the amount of dust in G4 suggests an efficient survival of dust (or inefficient feedback on dust) before and during initial quenching, efficient grain growth afterward, and perhaps a rejuvenation event during/after its quenching.
We performed the same calculation with other high-redshift galaxies (summarized in Figure~\ref{fig:dust_yield}), and two dust-rich quiescent galaxies (ALMA.14 and ZF-COS-19589) require similarly high dust yields.
Finally, as discussed in \citet{Michalowski2015}, it is also worth noting that Salpeter \citep{Salpeter1955} and top heavier IMF would require a comparable amount of and even higher dust yield; Chabrier IMF is the most conservative estimate.

\section{Conclusion}\label{sec:conclusion}
We reported the detection of cold dust in a massive quiescent galaxy ($\log({M_{\star}/M_{\odot}})\approx11$), G4, at $z\sim2$.
The galaxy is one of the six 2 mm continuum sources in the deep ALMA observations targeting a massive main-sequence galaxy at $z=2.21$.
We identified the optical/near-infrared counterparts for all but one galaxy.
This paper focused on one among them, G4, whose photometric redshift is estimated at $z\sim2$, close to the redshift of the Q2343-BX610 and G1 which are spectroscopically confirmed.

The avilable photometry suggest low SFR well below the $z=2$ main sequence ($\approx1.2$ dex) with $\log(SFR/M_{\star}) [yr^{-1}] \approx -10.2$.
The quiescent nature of the source is also supported by the $UVJ$ colors based on the best-fit SED.

We compiled the data in the literature for $z\gtrsim1$ quiescent galaxies and compared it with G4. 
We discovered the existence of a negative correlation between age and $\Delta$MS in quiescent galaxies at $z\sim1$, suggesting a passive evolution up to $z\sim1$.
Observational data at $z\sim2-3$ including G4 did not provide a strong indication of the negative correlation (and hence the passive evolution) with limited data points.
\textsc{simba} simulations showed a comparable negative relation at $z\sim1$ but at $z\sim2$ G4's age and $\Delta$MS is rare in their simulation box. 
More observational data is needed to determine whether G4 is unique (and rare) or whether the simulation underproduces such a population.

G4 exhibits unique features considering its estimated mass-weighted age ($\approx1-2$ Gyr), sSFR, and dust-to-stellar mass ratio compared to other dust-poor quiescent galaxies: the galaxy is dust-rich for its stellar mass ($\log({M_{\rm d}/M_{\star}}) = -2.71 \pm 0.26$) and its age (1-2 Gyr).
Based on the dust-to-stellar mass ratio and age relation, a longer dust depletion time is suggested, similar to dusty quiescent galaxies observed in local and intermediate redshift ($\tau=2.50-2.75$ Gyr).
This may indicate a potential evolutionary connection with each other.

The $UVJ$ color trajectories imply that the galaxy has experienced a fast-quenching mode.
However, morphological quenching along with inefficient (initial) feedback would be required to explain its dust-rich nature and G4's morphology supports this.
The high dust yield of G4 supports an inefficient quenching mode requiring efficient survival of dust and rejuvenation, perhaps assisted by efficient grain growth.\\

Our discovery encourages many observational programs for follow-up. 
Future rest-frame optical and high-frequency submillimeter observations to constrain the residual star-forming activity (if any), resolved distribution of stellar metallicity and age, stellar and gas kinematics, and deeper CO/\cone/\ctwo\, observations to constrain gas-to-dust ratio will give us more hints on the origin of the dust observed in the dust-rich quiescent galaxies and quenching mechanisms in the early universe.
Finding a non-negligible number of dust-rich galaxies (4/9) in the literature also pushes us to build a large sample of such, which will further help us understand the formation of these galaxies in the broader context of galaxy evolutionary studies and quenching in high redshift.

The deep ALMA Band 4 integration on a single field with a wealth of data allowed us to uncover a dust-rich quiescent candidate at $z=2$. 
This offered a new and complementary window to study high redshift quiescent galaxies and understand the connection between the available gas/dust and galaxy quenching. 
Based on this, we envisage deep ALMA observations of individual quiescent galaxies at high redshift (in a few selected areas of the sky) will also provide another insightful guidance to galaxy quenching at high redshift, together with statistical studies of the quiescent population from a large survey.

\section*{Acknowledgements}
We thank the referee for a careful reading of the manuscript and very helpful comments that helped to improve the paper. M.M. Lee thanks Francesco Valentino and Katherine E. Whitaker for general discussions on high redshift quiescent galaxies and their dust observations. M.M. Lee also thanks Steven Gilman for helpful comments on the data analysis, Aswin Vijayan for comments on cosmological simulations, M\'{e}d\'{e}ric Boquien for the comments on the CIGALE code, and David Bl\'{a}nquez-Ses\'{e} for offering the median stack $UVJ$ colors of $z\sim1$ quiescent galaxies.
This project has received funding from the European Union’s Horizon 2020 research and innovation programme under the Marie Skłodowska-Curie grant agreement No 101107795.
The Cosmic Dawn Center (DAWN) is funded by the Danish National Research Foundation under grant DNRF140.
This work was partially supported by DeiC National HPC (DeiC-DTU-L-20210103) and by grant AST2009278 from the US NSF (CCS). 
R.H.-C. thanks to the Max Planck Society for support under the Partner Group project "The Baryon Cycle in Galaxies" between the Max Planck for Extraterrestrial Physics and the Universidad de Concepción. R.H-C. also  gratefully acknowledges financial support from Millenium Nucleus NCN19058 
(TITANs), and ANID BASAL projects ACE210002 and FB210003.
This paper makes use of the following ALMA data: 
ADS/JAO.ALMA\#2019.1.01362.S, \#2019.1.00853.S,
\#2017.1.00856.S, \#2017.1.01045.S,
\#2015.1.00250.S,
\#2013.1.00059S. ALMA is a partnership of ESO (representing its member states), NSF (USA), and NINS (Japan), together with NRC (Canada), MOST and ASIAA (Taiwan), and KASI (Republic of Korea), in cooperation with the Republic of Chile. The Joint ALMA Observatory is operated by ESO, AUI/NRAO, and NAOJ.

\section*{Data Availability}
The processed ALMA map of G4, full SED fitting results, and a table of compiled quiescent galaxies are available upon request.




\bibliographystyle{mnras}
\bibliography{minjujournal_v2} 




\appendix


\appendix

\section{Photometric redshift}\label{app:sed}
In addition to \textsc{eazy-py} run, we explore two other different SED fitting codes, \textsc{bagpipes} \citep{Carnall2018} and \textsc{cigale} \citep{Boquien2019}, to cross-check the photometric constraints of G4.
The free parameters and assumptions except for SFH are the same as described in Section~\ref{sec:sed_model}.
For \textsc{Bagpipes}, we choose to use a double-power law for simultaneously fitting the photo-$z$, because the fit with delayed-$\tau$ and $\tau$ model were not well converged with high $\chi^2$ ($\gg30$). The inspection of the corner plots of the fitted parameters also convinced us to adopt the assumption of SFH to double-power law.

For \textsc{bagpipes}, we take Gaussian and uniform priors to fit the redshift. For the Gaussian priors, the center is set at the best photo-$z$ from \textsc{eazy-py} with relatively broad width (\texttt{redshift\_prior\_sigma}\,($\sigma_{\rm zprior})  = 2.0$). For \textsc{cigale}, we put uniform priors of redshift between 0 and 6.

\begin{figure}
\centering
\includegraphics[width=0.48\textwidth, bb=0 0 1850 1300]{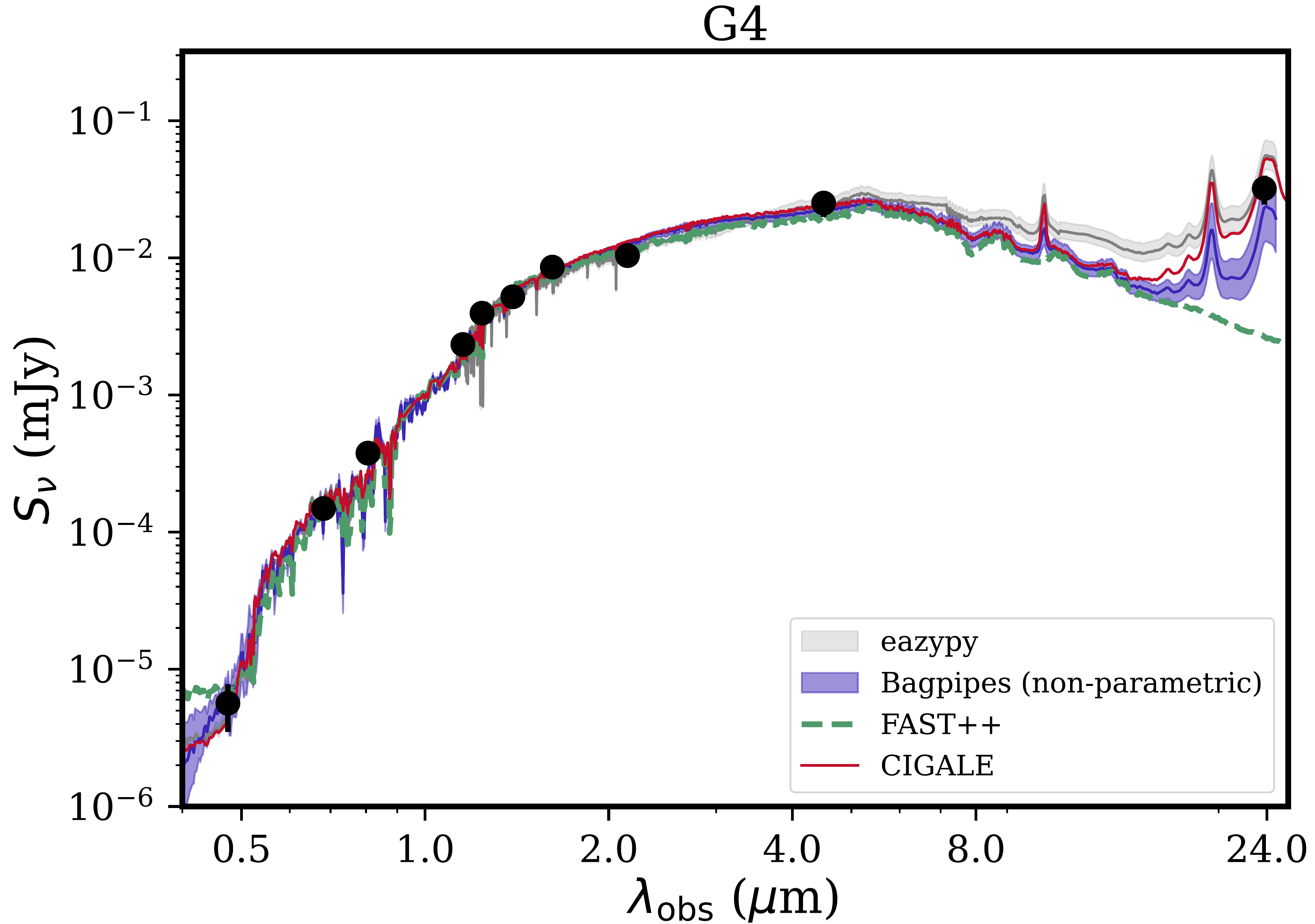}
\caption{The best-fit SED from \textsc{fast++}, \textsc{bagpipes} and \textsc{CIGALE} together with \textsc{eazy-py} fit by fixing redshift to $z=2.13$.\label{fig:sedbestfit}}
\end{figure}

Table~\ref{tab:sedzphot} shows the best fit without fixing the redshift using \textsc{bagpipes} and \textsc{cigale} for G4. 
All SED codes give a redshift solution of $z\sim2$ for G4.
The uniform prior from \textsc{bagpipes} did not provide a good convergence giving higher uncertainty of the $UVJ$ colors. 
We also note that at any redshift the estimated star-formation rate from \textsc{bagpipes} is close to zero, making the galaxy quiescent.
Finally, Figure~\ref{fig:sedbestfit} shows the best fit listed in Table~\ref{tab:sedbestfit} fixing the redshift at $z=2.13$ to illustrate that all SED codes fit the opt/NIR data points well.

\begin{table*}
\begin{center}
\caption{Best-fit parameters with redshift allowed to vary for G4\label{tab:sedzphot}}
\begin{tabular}{c|ccc}
\hline \hline
\multicolumn{4}{c}{G4} \\ \hline
&Bagpipes  & Bagpipes &  CIGALE \\ \hline
Redshift prior & 2.13 (2.0)$^{a}$ & [0,6]  & [0,6]\\ 
                 & Gaussian & uniform & uniform \\
Redshift		&	2.06 	&  2.07  &	$1.91\pm0.23$	\\
                & [2.02,2.10] & [0.54,2.11]\\
$\log M_{\star}$	& 11.14  &  11.10 &	$11.11\pm0.21$	\\
                 & [11.08,11.19]    &  [10.49,11.16]  &	\\
SFR$_{\rm SED}$$^{\dagger}$&		0.00$^{c}$    &  0.00 &	6e-4$\pm$5e-4$^{c,d}$	\\
		          & [0.00,0.00]	& [0.00,0.00]	&       \\
log($\Delta$MS)$^{\dagger}$	&	-99.66	   &  -185.13   &   -5.4 \\
$\log$(age)$^{b}$	&	8.88	 	&  8.89  & 	$9.26\pm0.32$	\\ 
			     &  [8.81,8.99] & [8.79,9.77]\\
SFH		&  double-power law  & double-power law & delayed-$\tau$	\\ 
$A_{\rm V}$ &     0.91      &     0.83   &      0.63$\pm$0.27\\ 
            &    [0.72,1.06]   &    [0.59,3.21]  \\ 
$U-V$ &     1.84$\pm$0.06   & 1.82$\pm$3.48 & $1.68\pm0.19$\\
$V-J$ &     1.14$\pm$0.11   & 1.19$\pm$1.53 & $1.07\pm0.25$\\ 
$\chi^{2 \ddagger}$    &      15.2    &      16.1  &   10.9   \\ \hline \hline 
\end{tabular}
    \begin{tablenotes}
      \small
     \item[$\dagger$] $^{\dagger}$: We note the limitation of the SFR from the SED which is likely to introduce an unrealistically low value of SFR and hence $\Delta$MS especially for the parametric SFH models. The photometry (\Rs-band) and the best-fit dust attenuation ($A_{\rm V}$) suggest the SFR of $\sim8\, M_{\odot}$ yr$^{-1}$ and $\log(\Delta$MS) $\sim -1.6$ (see also the main text in Section~\ref{sec:sedsfr}).
      \item[$\ddagger$]  $^{\ddagger}$: We quote the absolute $\chi^{2}$ value, as the templates employed (as is the case for many SED fitting codes) are not independent of each other, and degrees of freedom are ill-defined (e.g., \citealt{Smith2012}).

      \item[a] $^{a}$: The number in the parenthesis is the width of the Gaussian prior (\texttt{redshift\_prior\_sigma}). The redshift was then set to vary between 0.1 and 12. Based on this setup, the redshift space has a hard limit at 3 sigma, meaning that $z=(0,8.13)$ is explored.
      \item[b] $^{b}$: mass-weighted.
      \item[c] $^{c}$: averaged over 100 Myr
      \item[d] $^{d}$: including the ALMA 2~mm data point.
      \end{tablenotes}
\end{center}
\end{table*}

\section{A summary table of quiescent galaxies from the literature}\label{app:qg}
We provide a summary of $z>1$ quiescent galaxies with dust/gas constraints used in this paper in Table~\ref{tab:otherQG} including the work presented here. The corresponding references are also listed.
\begin{table*}
\caption{Compilation of quiescent galaxies at $z>1$ studied with dust, CO and/or [CI] observations \label{tab:otherQG}}
\begin{tabular}{lcccccccccl}
 \hline  \hline
             Source &  Redshift &     $\mu$ &   $S_{\nu}$ &  Freq  &     $I_{\rm line}$ &   $\log(\mu M_{\star}/M_{\odot})$ &  $\mu$SFR &  Age  &    $A_{V}$ &                                       Ref. \\ 
                    &           &        &  ($\mu$Jy)     & (GHz)   & (Jy km s$^{-1}$) &  & ($M_{\odot}$ yr$^{-1})$     & (Gyr)\\ 
            (1) & (2) & (3) & (4) & (5) & (6) & (7) & (8) & (9) & (10) & (11)     \\ \hline
          EGS-18045 &     1.012 &  1.0 & - & - &    $0.88 \pm 0.31^{a}$ &  11.33 &           15.8$^{*}$ &            3.80 &    0.84 &                         [1,2]\\
          EGS-20106 &     1.062 &  1.0 & - & - &    $0.43\pm 0.06^{b}$ &  11.25 &           26.9$^{*}$ &            2.60 &    0.93 &                         [1,2] \\
              22260 &     1.240 &  1.0 & -& - &    $0.09 \pm 0.019^{b}$ &  11.51 &            5.3$^{**}$ &            4.60 & - &                   [3,4] \\
          EGS-17533 &     1.264 &  1.0 & -& - &    $0.060\pm 0.014^{b}$ &  10.78 &           14.1$^{*}$ &            1.00 &    0.59 &                         [1,5] \\
              34879 &     1.322 &  1.0 &  & - &    $<0.041^{b,f}$ &  11.32 &           22.9$^{*}$ &            2.10 & - &                      [3,6] \\
             217431 &     1.428 &  1.0 & - & - &    $<0.32^{b,f}$ &  11.57 &            0.4$^{**}$ &            3.98 &    0.25 &                     [7,8] \\
             307881 &     1.429 &  1.0 & - & - &    $<0.057^{b,f}$ &  11.63 &            5.0$^{*}$ &            3.20 & - &                    [3,8]\\
            ALMA.14 &     1.451 &  1.0 & $1050 \pm 240$ &  344.8 &    $0.227 \pm 0.040^{b}$ &  10.96 &            3.0$^{**}$ &          - & - &                     [9,10] \\
              20866 &     1.522 &  1.0 & - & - &    $<0.071^{b,f}$ &  11.46 &           12.8$^{*}$ &            1.70 & - &                   [3,4]  \\
              21434 &     1.522 &  1.0 & - & - &    $<0.104^{b,f}$ &  11.39 &           19.1$^{*}$ &            1.20 & - &       [3,4,11]  \\
              34265 &     1.582 &  1.0 & - & - &   $<0.053^{b,f}$ &  11.51 &            7.4$^{*}$ &            1.30 & - &                      [3,6] \\
           MRG-1341 &     1.594 & 30.0 &   $<27^{f}$ &  232.7 & - &  11.62 &            3.0$^{**}$ &            1.19 &    0.40 &    [12,13,14,15] \\
              30169 &     1.629 &  1.0 & - & - &    $0.060 \pm  0.010^{c,f}$ &  11.22 &           12.0$^{**}$ &          - & - &                               [23] \\
          MRG-S0851 &     1.880 &  9.6 &  $<720^{f}$ &  272.0 & - &  12.00 &            57.6$^{**}$ &          - & - &                     [15,16]  \\
           MRG-0138 &     1.949 & 12.5 & $285\pm29$ &  232.7 & - &  12.77 &           31.0$^{**}$ &            1.35 &    0.35 & [12,13,16,17] \\
           G4 & 2.13$^{\dagger}$ & 1.0 & $44\pm12$ & 140.0 & - & 11.22 &             0.01$^{**}$ &          1.86 & 0.51 & This work \\
           MRG-2129 &     2.149 &  4.6 &  $164\pm17$ &  232.7 &    $<0.016^{d,f}$ &  11.62 &   0.12$^{**}$ &            0.61 &    1.10 &  [12,13,14,18] \\
           MRG-0150 &     2.635 &  4.4 &   $<52^{f}$ &  232.7 &  - &  12.06 &           $<22.0^{**}$ &            0.76 &    0.61 &            [12,13,17] \\
           Galaxy-D &     2.900 &  1.0 &   $<84^{f}$ &  344.0 & - &  11.00 &            $<4.0^{**}$&            1.60 &    0.10 &                                 [19] \\
           MRG-0454 &     2.922 & 10.9 &   $<42^{f}$ &  232.7 & - &  11.65 &           68.8$^{**}$ &            0.33 &    0.90 &    [12,13,14,15]\\
           MRG-1423 &     3.209 &  2.7 &   $28\pm11$ &  232.7 &  - &  11.01 &          428.0$^{**}$ &            0.12 &    0.80 &               [12,13,14,15]  \\
        ZF-UDS-8197 &     3.543 &  1.0 &  $<100^{f}$ &  344.8 &    $<0.040^{e,f}$ &  10.56 &  4.0$^{***}$ &          0.95 &    0.00 &                    [20,21] \\
       ZF-COS-19589 &     3.715$^{\ddagger}$ &  1.0 &  $210\pm40$&  344.8 &    $<0.060^{e,f}$ &  10.79 &           33.0$^{*}$ &            0.53 &    0.90 &                    [20,21] \\
        ZF-COS-20115 &     3.715 &  1.0 &   $<50^{f}$ &  302.4 &    $<0.110^{e,f}$ &  11.06 &   0.0$^{**}$ &          0.76 &    0.30 &      [20,21,22] \\
       ZF-COS-18842 &     3.782$^{\ddagger}$ &  1.0 &  $<130^{f}$ &  344.8 &    $<0.050^{e,f}$ &  10.65 &           $3.6^{***}$ &          0.54 &    0.00 &                    [20,21] \\ \hline
\end{tabular}
    \begin{tablenotes}
      \small
        \item Columns: (1) ID (2) Redshift (spectroscopic redshift, if it is not specifically noted), (3) Magnification factor. If $\mu>1.0$, it is lensed. (4) Observed dust continuum flux. (5) Observed frequency. (6) Observed integrated line flux (7) Stellar mass in log (without lensing correction) (8) SFR (without lensing correction). (9) Mass-weighted stellar age. (10) Dust attenuation in $V$-band magnitude. (11) References.
        \item References : [1] \citet{Belli2021}, [2] \citet{Belli2019}, [3] \citet{Williams2021}, [4] \citet{Bezanson2013}, [5] \citet{Suess2017}, [6] \citet{Belli2015}, [7] \citet{Sargent2015}, [8] \citet{Onodera2012}, [9] \citet{Hayashi2017}, [10] \citet{Hayashi2018}, [11] \citet{Bezanson2019}, [12] \citet{Gobat2022}, [13] \citet{Whitaker2021}, [14] \citet{Man2021}, [15] \citet{Akhshik2023} [16] \citet{Caliendo2021}, [17] \citet{Newman2018a}, [18] \citet{Morishita2022}, [19] \citet{Kalita2021}, [20] \citet{Suzuki2022}, [21] \citet{Schreiber2018b},  [22] \citet{Glazebrook2017} , [23] \citet{Rudnick2017}
        \item Notes: For MRG-1341, MRG-0138, MRG-2129, MRG-0150, MRG-0454, and MRG-1423, we took dust flux from Table 1 in \citet{Gobat2022} for the point-source model, which would be comparable to the original study of \citet{Whitaker2021}. For MRG-2129, \citet{Morishita2022} measured the flux in ALMA Band 6 (1.2 mm, Table 2 therein) which is consistent within the error bar. As discussed in \citet{Gobat2022}, if dust is extended, the flux (upper limit) can become higher by up a factor of 1.5 ($\sim 6$). The SFR values listed here are the maximum values obtained by different tracers (UV+IR, SED, [OII]/H$\beta$) available in the literature. For the stellar mass of these lensed sources is based on \citet{Newman2018a, Man2021} rather than reported in \citet{Gobat2022}. \citet{Man2021} adopted \citet{Kroupa2001} IMF for MRG-1341, MRG-0454, and MRG-1423 and we did not make additional changes for Chabrier IMF as the difference is expected to be small ($\approx$0.03 dex) according to \citet{Madau2014}.
        \item $^{a}$ : \cothree, $^{b}$ : \cotwo, $^{c}$ : \coone, $^{d}$ : \citwo, $^{e}$ : \cione
        \item $^{f}$ : 3$\sigma$ upper limit
        \item $^{*}$ : SFR based on UV+IR or IR-only
        \item $^{**}$ : SFR based on SED fitting
        \item $^{***}$ : SFR based on H$\beta$
        \item $^{\dagger}$ : photometric redshift
        \item $^{\ddagger}$ : uncertain redshift
      \end{tablenotes}

\end{table*}


\bsp	
\label{lastpage}
\end{document}